\documentclass[prd,twocolumn,superscriptaddress,floatfix,nofootinbib]{revtex4-2}
\pdfoutput=1 
\usepackage[T1]{fontenc} 
\usepackage{amsmath,amssymb,amsfonts,amsthm}
\usepackage{graphicx}
\usepackage[usenames,dvipsnames]{color}
\usepackage{enumitem}
\usepackage{verbatim}
\usepackage[percent]{overpic}
\usepackage{rotating}
\usepackage{hyperref}
\usepackage{array}
\usepackage{mathtools}
\usepackage{lmodern}
\usepackage[normalem]{ulem}
\usepackage{braket}
\usepackage{tensor}
\usepackage{dsfont}
\usepackage{bm}
\usepackage{tikz}

\usepackage{mathrsfs}

\usetikzlibrary{decorations.pathmorphing}
\usepackage{CJKutf8}

\usetikzlibrary{positioning}
\usetikzlibrary{calc,through,backgrounds}

\theoremstyle{definition}
\newtheorem{definition}{Definition}[section]
\newtheorem{theorem}{Theorem}[section]

\newcommand{\kako}[1]{\left( #1 \right)}
\newcommand{\kagikako}[1]{\left[ #1 \right]}
\newcommand{\ts}[1]{ _{\text{#1}} }

\allowdisplaybreaks

\DeclareMathOperator{\Tr}{Tr}

\newcommand{\dd}{\text{d}}

\newcommand{\id}{\mathds{1}}
\newcommand{\sx}{\mathsf{x}}

\newcommand{\ii}{\mathsf{i}}

\newcommand{\AAA}{\text{A}}
\newcommand{\BB}{\text{B}}

\begin{document}

\title{Harvesting mutual information from BTZ black hole spacetime}


\author{Kendra Bueley}
\email{kbueley@uwaterloo.ca}
\affiliation{Department of Physics and Astronomy, University of Waterloo, Waterloo, Ontario, N2L 3G1, Canada}

\author{Luosi Huang}
\email{luosi.huang@uwaterloo.ca}
\affiliation{Department of Physics and Astronomy, University of Waterloo, Waterloo, Ontario, N2L 3G1, Canada}

\author{Kensuke Gallock-Yoshimura}
\email{kgallock@uwaterloo.ca} 

\affiliation{Department of Physics and Astronomy, University of Waterloo, Waterloo, Ontario, N2L 3G1, Canada}


\author{Robert B. Mann}
\email{rbmann@uwaterloo.ca}
\affiliation{Department of Physics and Astronomy, University of Waterloo, Waterloo, Ontario, N2L 3G1, Canada}

\begin{abstract}
We investigate the correlation harvesting protocol for mutual information between two Unruh-DeWitt detectors in a static BTZ black hole spacetime. 
Here, the effects coming from communication and change in proper separation of the detectors are set to be negligible so that only a black hole affects the extracted mutual information. 
We find that, unlike the entanglement harvesting scenario, harvested mutual information is zero only when a detector reaches an event horizon, and that   although the Hawking effect and gravitational redshift both affect the extraction of mutual information, it is extreme Hawking radiation that inhibits the detectors from harvesting. 
\end{abstract}

\maketitle
\flushbottom

\section{Introduction}
Relativistic quantum information theory studies the interplay between quantum information processing and relativistic effects, and  has attracted much attention in   recent years. 
Many of the studies use either the Bogoliubov transformation or particle detector models, such as  Unruh-DeWitt (UDW) detectors \cite{Unruh1979evaporation, DeWitt1979}, to examine, for example, the quantum teleportation protocol \cite{PhysRevLett.91.180404, Landulfo2009suddendeath}, relativistic quantum communication channels \cite{Cliche2010channel, jonsson2017quantum, Landulfo2016magnus1, Simidzija2020transmit}, and entanglement degradation \cite{FuentesAliceFalls, AlsingDiracFields}. 

In particular the so-called \textit{entanglement harvesting protocol} is an operation where multiple initially uncorrelated particle detectors extract entanglement from the vacuum of a quantum field via local interaction \cite{Valentini1991nonlocalcorr, reznik2003entanglement, reznik2005violating}. 
Entanglement harvesting exploits the fact that vacuum states of a quantum field are entangled \cite{summers1985bell, summers1987bell}, and so even causally disconnected detectors can be correlated. 
Extensive research shows that the extracted entanglement is sensitive to spacetime properties   \cite{pozas2015harvesting, smith2016topology, kukita2017harvesting, henderson2018harvesting, ng2018AdS, cong2020horizon, FinnShockwave} and the state of motion of the detectors  \cite{salton2015acceleration,FooSuperpositionTrajectory,Liu:2021dnl}.

Our particular interest lies in the impact of black holes on harvested correlations between detectors. 
The first investigation of the harvesting protocol in a black hole spacetime was done by Henderson \textit{et al.} \cite{henderson2018harvesting}. 
It was shown that there is an \textit{entanglement shadow} --- a region   where two static detectors hovering outside a Ba\~{n}ados-Teitelboim-Zanelli (BTZ) black hole cannot harvest entanglement. 
This region is located near the black hole at sufficiently large angular separations of the detectors relative to the origin; it arises due to a combination of increased local Hawking temperature near the black hole and
gravitational redshift  suppression of nonlocal correlations.  
Other types of black holes have also been examined, including  static detectors in Schwarzschild and Vaidya spacetimes \cite{Tjoa2020vaidya}, topological black holes \cite{Campos-hyperbolicBH},
geons \cite{Henderson:2022oyd}, 
corotating detectors around a rotating BTZ black hole \cite{robbins2020entanglement}, and freely falling detectors in Schwarzschild spacetime \cite{Ken.Freefall.PhysRevD.104.025001}.

While most studies of the harvesting protocol focus on extraction of entanglement, little has been done for other correlations such as mutual information. 
Quantum mutual information quantifies the total amount of classical and quantum correlations including entanglement, and
a few investigations of mutual information harvesting have been carried  out~\cite{simidzija2018harvesting, GallockEntangledDetectors, SahuSabotage, pozas2015harvesting}.
Unlike entanglement, mutual information does not vanish anywhere outside the event horizon,  at least in $(1+1)$-dimensional settings
\cite{Tjoa2020vaidya, Ken.Freefall.PhysRevD.104.025001}. 
However a detailed study of how  Hawking radiation and gravitational redshift influence the harvesting protocol for mutual information has not yet been carried out. 

Our purpose here is to address this issue by examining the harvesting protocol for mutual information in the presence of a black hole. For simplicity, we consider the nonrotating BTZ black hole spacetime and,  as with other studies \cite{henderson2018harvesting}, 
consider identical UDW detectors that are static   outside of the horizon and do not communicate with each other \cite{TjoaSignal}. 
This setup allows us to  investigate  effects that are purely due to the black hole, such as redshift and the Hawking effect. 
We obtain two key results. First,
 although the amount of harvested correlation is reduced by these effects as detectors get close to a black hole and vanishes at the event horizon, the ``shadow'' for mutual information is absent. 
Since entanglement cannot be harvested near a black hole, the extracted correlation is either classical   or  nondistillable entanglement. 
Second, the death of mutual information at the event horizon is due to Hawking radiation and not gravitational redshift.

Our paper is organized as follows. 
In Sec. \ref{subsec:BTZ spacetime} we introduce the BTZ spacetime and the conformally coupled scalar field defined on this background. 
The final density matrix of UDW detectors after they interact with the quantum field is given in Sec. \ref{subsec:UDW detectors}. 
The mutual information can be written in terms of elements in this density matrix. 
Then we show how mutual information is affected by the black hole in Sec. \ref{sec: results}, followed by conclusion in Sec. \ref{sec: conclusion}. 
Throughout this paper, we use the units $\hbar = c=1$ and the signature $(-,+,+)$. 
Also, $\sx\coloneqq x^\mu$ denotes an event in  coordinates $x^\mu$.

\section{UDW detectors in BTZ spacetime}
\label{sec:UDW detectors in BTZ spacetime}

\subsection{BTZ spacetime and quantum fields}
\label{subsec:BTZ spacetime}

The BTZ spacetime \cite{BTZ1, BTZ2} is a $(2+1)$-dimensional black hole spacetime with a negative constant curvature. Its line-element in   static coordinates is given by
\begin{subequations}
\begin{align}
    \dd s^2
    &=
        - f(r) \dd t^2
        + \dfrac{ \dd r^2 }{f(r)}
        + r^2 
        \dd \phi^2\,,\\
    f(r)
    &=
        \dfrac{r^2}{\ell^2}  -M\,,
\end{align} 
\end{subequations}
with $t\in \mathbb{R}, r\in (0,\infty)$, and $\phi \in [0, 2\pi)$. 
The dimensionless quantity $M$ is the mass of the BTZ black hole and $\ell$ is known as the AdS length, which determines the negative curvature of the spacetime. 
The event horizon $r\ts{h}$ of a nonrotating BTZ black hole can be obtained from the roots of $f(r)=0$, which is 
\begin{align}
    r\ts{h}
    &=
        \ell \sqrt{M}\,. 
\end{align}

We will examine correlations between two detectors in terms of their proper distance. 
For $r_2 > r_1 \geq r\ts{h}$, the proper distance, $d(r_1, r_2)$, between two spacetime points $(t,r_1,\phi)$ and $(t,r_2,\phi)$ is given by 
\begin{align}
    d(r_1, r_2)
    &=
        \ell 
        \ln 
        \kako{
            \dfrac{r_2 + \sqrt{ r_2^2 - r\ts{h}^2 }}
            {r_1 + \sqrt{ r_1^2 - r\ts{h}^2 }}
        }. \label{eq:proper distance}
\end{align}
We will  consider the case where detector A is closer to the event horizon, namely $r\ts{B}>r\ts{A}>r\ts{h}$, and fix the proper separation between the detectors $d\ts{AB}\coloneqq d(r\ts{A}, r\ts{B})$. 
We also write $d_j\coloneqq d(r\ts{h}, r_j)$, $j\in \{ \AAA, \BB \}$ for simplicity.

The Hawking temperature $T\ts{H}$ of a BTZ black hole is known to be\footnote{Note that $T\ts{H} \propto \sqrt{M}$ in contrast to the
$T\ts{H} \propto 1/{M}$ behavior for a Schwarzschild black hole. } $T\ts{H}=r\ts{h}/2\pi \ell^2$ and the local temperature $T_j$ at radial position $r=r_j$ is given by 
\begin{align}
    T_j
    &=
        \dfrac{T\ts{H}}{ \gamma_j }\,, \label{eq:local temperature}
\end{align}
where 
\begin{align}
    \gamma_j
    &=
        \dfrac{ \sqrt{ r_j^2 - r\ts{h}^2 } }{ \ell }
        ~~~~~(r_j\geq r\ts{h}) \label{eq:redshift}
\end{align}
is the redshift factor. 
For $r\ts{B}>r\ts{A}>r\ts{h}$, the redshift factors for detectors A and B can be written in terms of the proper distances: 
\begin{subequations}
\begin{align}
    \gamma\ts{A}
    &=
        \dfrac{r\ts{h}}{\ell} \sinh \dfrac{d\ts{A}}{\ell}\,, \\
    \gamma\ts{B}
    &=
        \dfrac{r\ts{h}}{\ell} \sinh \dfrac{d\ts{AB} + d\ts{A}}{ \ell }\,.
\end{align}
\label{eq:redshift and distance}
\end{subequations}
More precisely, the temperature \eqref{eq:local temperature} is known as the Kubo-Martin-Schwinger (KMS) temperature at $r=r_j$.

We now introduce a conformally coupled quantum scalar field $\hat \phi(\sx)$ in the BTZ spacetime. 
Choosing the Hartle-Hawking vacuum $\ket{0}$, the Wightman function $W\ts{BTZ}(\sx, \sx')\coloneqq \bra{0} \hat \phi(\sx) \hat \phi(\sx') \ket{0}$ can be constructed from the image sum of the Wightman function in the AdS$_3$ spacetime, $W\ts{AdS}(\sx,\sx')$ \cite{LifschytzBTZ}. 
Let $\Gamma:(t,r,\phi) \to (t,r,\phi+2\pi)$ represent the identification of a point $\sx$ in AdS$_3$ spacetime. 
Then $W\ts{BTZ}$ is known to be 
\begin{align}
    &W\ts{BTZ}(\sx, \sx')
    =
        \sum_{n=-\infty}^\infty
        W\ts{AdS}(\sx, \Gamma^n \sx') \notag \\
    &=
        \dfrac{1}{ 4\pi \sqrt{2}\ell }
        \sum_{n=-\infty}^\infty
        \kagikako{
            \dfrac{1}{ \sqrt{ \sigma_\epsilon (\sx, \Gamma^n \sx') } }
            -
            \dfrac{\zeta}{ \sqrt{ \sigma_\epsilon (\sx, \Gamma^n \sx')+2 } }
        }, \label{eq:BTZ Wightman}
\end{align}
where 
\begin{align}
    \sigma_\epsilon (\sx, \Gamma^n \sx')
    &= 
        \dfrac{r r' }{ r\ts{h}^2} 
        \cosh 
        \kagikako{
            \dfrac{r\ts{h}}{\ell} (\Delta \phi - 2\pi n )
        }
        -1 \label{sigma} \\
        &-\dfrac{ \sqrt{ (r^2-r\ts{h}^2) (r^{\prime 2}-r\ts{h}^2) } }{ r\ts{h}^2 } 
        \cosh 
        \kako{
            \dfrac{r\ts{h}}{\ell^2} 
            \Delta t - \ii \epsilon
\nonumber        }
\end{align}
with $\Delta \phi\coloneqq \phi-\phi'$, $\Delta t\coloneqq t-t'$, and $\epsilon$ is the ultra-violet regulator. 
$\zeta \in \{ -1, 0, 1 \}$ specifies the boundary condition for the field at the spatial infinity: Neumann $\zeta=-1$, transparent $\zeta=0$, and Dirichlet $\zeta=1$. 
We will choose $\zeta=1$ throughout this article. 
We also note that $n=0$ term in the Wightman function is called the AdS-Rindler term
\cite{Henderson2019anti-hawking,Campos-RobinBC,Robbins-Anti-Hawking}
, which corresponds to a uniformly accelerating detector in AdS$_3$ spacetime
\cite{Jennings:2010vk};
the remaining ($n\neq 0$) terms are known as the BTZ terms, which yield the black hole contribution.

\subsection{UDW detectors}
\label{subsec:UDW detectors}
A UDW detector is a two-level quantum system consisting of  ground and excited states with  energy gap $\Omega$, and   can be considered as a qubit. 
Let us prepare two pointlike UDW detectors A and B. 
Each of them has their own proper time $\tau_j\,(j\in \{ \AAA, \BB \})$ and will interact with the local quantum field $\hat \phi(\sx)$ through the following interaction Hamiltonian 
\begin{align}
    \hat H_j^{ \tau_j } ( \tau_j )
    &=
        \lambda_j \chi_j(\tau_j) \hat \mu_j(\tau_j) 
        \otimes \hat \phi(\sx_j(\tau_j)),\,j\in \{ \AAA, \BB \}
\end{align}
in the interaction picture, where $\lambda_j$ is the coupling strength and $\chi_j(\tau_j)$ is the switching function describing the interaction duration of detector-$j$. 
The operator $\hat \mu_j(\tau_j) $ is the so-called monopole moment, which describes the dynamics of a detector and  is given by 
\begin{align}
    \hat \mu_j(\tau_j) 
    &=
        \ket{e_j} \bra{g_j} e^{ \ii \Omega_j \tau_j }
        +
        \ket{g_j} \bra{e_j} e^{ -\ii \Omega_j \tau_j },
\end{align}
where $\ket{g_j}, \ket{e_j}$, and $\Omega_j$ are respectively the ground, excited states of detector-$j$ and the energy gap between them. 
$\hat \phi(\sx_j(\tau_j))$ is the pullback of the field operator along detector-$j$'s trajectory. 
We put the superscript on the Hamiltonian $\hat H_j^{ \tau_j } ( \tau_j )$ to indicate that it is the generator of time-translation with respect to the proper time $\tau_j$. 
In what follows, we assume that the detectors have the same coupling strength $\lambda$ and energy gap $\Omega$. 

Let us write the total interaction Hamiltonian as a generator of time-translation with respect to the common time $t$ in the BTZ spacetime: 
\begin{align}
    \hat H\ts{I}^t(t)
    &=
        \dfrac{\dd \tau\ts{A}}{\dd t} 
        \hat H\ts{A}^{ \tau\ts{A} }\big( \tau\ts{A}(t) \big)
        +
        \dfrac{\dd \tau\ts{B}}{\dd t} 
        \hat H\ts{B}^{ \tau\ts{B} }\big( \tau\ts{B}(t) \big) \,, 
\end{align}
so that the time-evolution operator $\hat U\ts{I}$ is given by a time-ordering symbol $\mathcal{T}_t$ with respect to $t$ \cite{EMM.Relativistic.quantum.optics,Tales2020GRQO}: 
\begin{align}
    \hat U\ts{I}
    &=
        \mathcal{T}_t 
        \exp 
        \kako{
            -\ii \int_{\mathbb{R}} \dd t\,\hat H\ts{I}^t(t)
        } .
\end{align}

We now assume that the coupling strength $\lambda$ is small and apply the Dyson series expansion to $\hat U\ts{I}$: 
\begin{subequations}
\begin{align}
    \hat U\ts{I}
    &=
        \id + \hat U^{(1)} + \hat U^{(2)} + \mathcal{O}(\lambda^3)\,,\\
    \hat U^{(1)}
    &=
        -\ii \int_{-\infty}^\infty \dd t\,\hat H\ts{I}^t(t)\,,\\
    \hat U^{(2)}
    &=
        - \int_{-\infty}^\infty \dd t_1
        \int_{-\infty}^{t_1} \dd t_2\,
        \hat H\ts{I}^t(t_1) \hat H\ts{I}^t(t_2)\,.
\end{align}
\end{subequations}
We further assume that the detectors and the field are initially in their ground states and uncorrelated. 
The initial state $\rho_0$ of the total system is then
\begin{align}
    \rho_0
    &=
        \ket{g\ts{A}} \bra{g\ts{A}}
        \otimes 
        \ket{g\ts{B}} \bra{g\ts{B}}
        \otimes 
        \ket{0}\bra{0}\,,
\end{align}
where $\ket{0}$ is the vacuum state of the field. 
After the interaction, the final total density matrix $\rho\ts{tot}$ reads
\begin{align}
    \rho\ts{tot}
    &=
        \hat U\ts{I} \rho_0 \hat U\ts{I}^\dag \notag \\
    &=
        \rho_0 
        + 
        \rho^{(1,1)}
        +
        \rho^{(2,0)}
        +
        \rho^{(0,2)}
        +
        \mathcal{O}(\lambda^4),
\end{align}
where $\rho^{(i,j)}=\hat U^{(i)} \rho_0 \hat U^{(j)\dagger}$ and we used the fact that all the odd-power terms of $\lambda$ vanish \cite{pozas2015harvesting}.

The final density matrix $\rho\ts{AB}=\Tr_\phi[\rho\ts{tot}]$ of the detectors in the basis $\{ \ket{g\ts{A} g\ts{B}} , \ket{g\ts{A} e\ts{B}}, \ket{e\ts{A} g\ts{B}}, \ket{e\ts{A} e\ts{B}} \}$ is known to be
\begin{align}
    \rho\ts{AB}
    &=
        \left[
        \begin{array}{cccc}
        1-\mathcal{L}\ts{AA}-\mathcal{L}\ts{BB} &0 &0 &\mathcal{M}^*  \\
        0 &\mathcal{L}\ts{BB} &\mathcal{L}\ts{AB}^* &0  \\
        0 &\mathcal{L}\ts{AB} &\mathcal{L}\ts{AA} &0  \\
        \mathcal{M} &0 &0 &0 
        \end{array}
        \right]
        + \mathcal{O}(\lambda^4)\,,
\end{align}
where
\begin{align}
    \mathcal{L}_{ij}
    &=
        \lambda^2
        \int_{\mathbb{R}} \dd \tau_i
        \int_{\mathbb{R}} \dd \tau_j'\,
        \chi_i(\tau_i) \chi_j(\tau_j')
        e^{ -\ii \Omega (\tau_i - \tau_j') } \notag \\
        &\qquad\qquad\qquad \qquad\times 
        W\ts{BTZ}\big( \sx_i(\tau_i), \sx_j(\tau_j') \big)\,, \\
    \mathcal{M}
    &=
        -\lambda^2
        \int_{\mathbb{R}} \dd \tau\ts{A}
        \int_{\mathbb{R}} \dd \tau\ts{B}\,
        \chi\ts{A}(\tau\ts{A}) \chi\ts{B}(\tau\ts{B})
        e^{ \ii \Omega (\tau\ts{A} + \tau\ts{B}) } \notag \\
        &\hspace{5mm}\times 
        \big[ 
            \Theta \big( t(\tau\ts{A}) - t(\tau\ts{B}) \big)
            W\ts{BTZ} \big( \sx\ts{A}(\tau\ts{A}), \sx\ts{B}(\tau\ts{B}) \big) \notag \\
            &\hspace{1cm}
            +
            \Theta \big( t(\tau\ts{B}) - t(\tau\ts{A}) \big)
            W\ts{BTZ} \big( \sx\ts{B}(\tau\ts{B}), \sx\ts{A}(\tau\ts{A}) \big)
        \big]\,,
\end{align}
where $\Theta(t)$ is the Heaviside step function. 
The off-diagonal elements $\mathcal{M}$ and $\mathcal{L}\ts{AB}$ are responsible for entanglement and mutual information, respectively. 
It is worth knowing that the elements satisfy $\mathcal{L}\ts{AA}\mathcal{L}\ts{BB}\geq |\mathcal{L}\ts{AB}|^2$ \cite{smith2016topology}.

The total classical and quantum correlations are
given by 
\begin{align}
    I\ts{AB}
    &=
        \mathcal{L}_+ \ln \mathcal{L}_+
        + 
        \mathcal{L}_- \ln \mathcal{L}_- \notag \\
        &\hspace{5mm}-
        \mathcal{L}\ts{AA} \ln \mathcal{L}\ts{AA}
        -
        \mathcal{L}\ts{BB} \ln \mathcal{L}\ts{BB}
        + \mathcal{O}(\lambda^4)
        \,,
\end{align}
which is the  mutual information $I\ts{AB}$ between detectors A and B, where 
\begin{align}
    \mathcal{L}_\pm
    &\coloneqq
        \dfrac{1}{2}
        \kako{
            \mathcal{L}\ts{AA}
            +
            \mathcal{L}\ts{BB}
            \pm 
            \sqrt{ (\mathcal{L}\ts{AA}-\mathcal{L}\ts{BB})^2 + 4 |\mathcal{L}\ts{AB}|^2 }
        }.
\end{align}
Note that $I\ts{AB}=0$ if $\mathcal{L}\ts{AB}=0$, and from the condition, $\mathcal{L}\ts{AA}\mathcal{L}\ts{BB}\geq |\mathcal{L}\ts{AB}|^2$, we have $\mathcal{L}\ts{AA}\mathcal{L}\ts{BB}=0 \Rightarrow |\mathcal{L}\ts{AB}|=0 \Rightarrow I\ts{AB}=0$. 

We will use a Gaussian switching function with a typical interaction duration $\sigma$,
\begin{align}
    \chi_j(\tau_j)
    &=
        e^{ -\tau_j^2/2\sigma^2 },
\end{align}
in which case  $\mathcal{L}_{ij}$ can be reduced to single integrals as shown in Appendix \ref{app:Derivation of Lij}.

\begin{figure*}[tp]
    \centering
    \includegraphics[width=\linewidth]{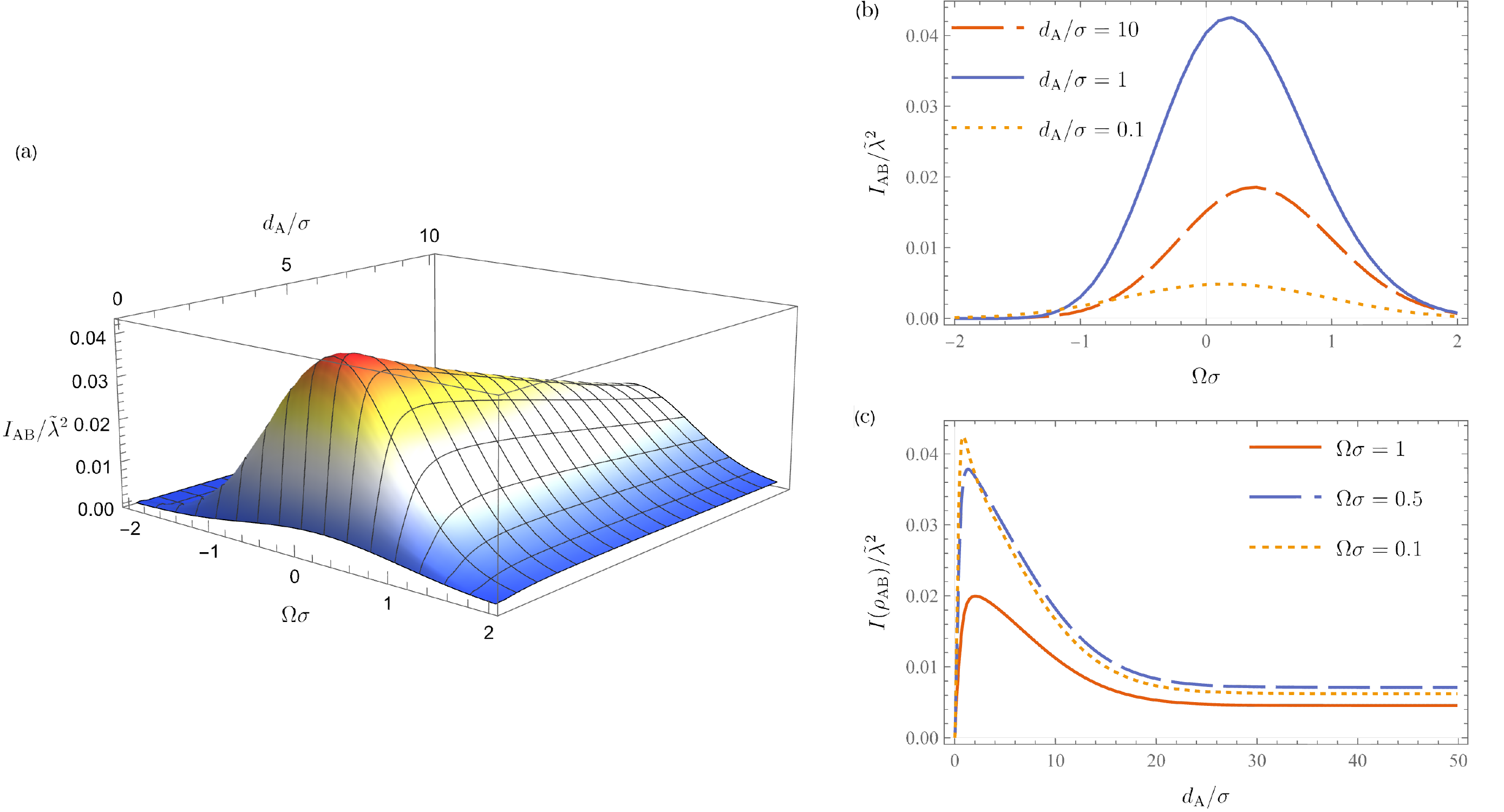}\\
    \caption{
    ~(a) 3D plot of mutual information $I\ts{AB}/\tilde{\lambda}^2$ as a function of energy gap $\Omega \sigma$ and proper distance $d\ts{A}/\sigma$. 
    Here, $\ell/\sigma=10, M=10^{-2}$, and $d\ts{AB}/\sigma=7$. 
    (b) Some slices of (a) with constant $d\ts{A}/\sigma=10,1$, and 0.1. 
    (c) Slices of (a) with constant $\Omega \sigma=1, 0.5$, and 0.1. }
    \label{fig:MIOmegadhaplots}
\end{figure*}

\section{Results}
\label{sec: results}
In this section we investigate how a black hole affects the extraction of mutual information. 
Detectors A and B are both static outside the black hole and aligned along an axis through its center
[so that $\Delta\phi = 0$ in \eqref{sigma}], and they switch at the same time. 
To see the effects purely coming from the black hole, we always fix the proper separation, $d\ts{AB}$, between A and B in such a way that the contribution from communication is negligible. 
In what follows we write quantities in units of the Gaussian width $\sigma$. 
For example, $\tilde{\lambda}\coloneqq \lambda \sqrt{\sigma}$ denotes a dimensionless coupling constant.

Let us first analyze the dependence on the energy gap $\Omega$ and the proper distance $d\ts{A}$ between detector A and the event horizon in Fig.~\ref{fig:MIOmegadhaplots}. 
We fix $\ell/\sigma=10, M=10^{-2}$ and $d\ts{AB}/\sigma=7$, and treat mutual information as a function of $\Omega$ and $d\ts{A}$: $I\ts{AB}=I\ts{AB}(\Omega, d\ts{A})$. 
Figures~\ref{fig:MIOmegadhaplots}(b) and (c) depict slices of (a) with constant $d\ts{A}/\sigma$ and $\Omega \sigma$, respectively. 
From Figs. \ref{fig:MIOmegadhaplots}(a) and (b), one finds that there exists an optimal value of $\Omega$ to extract correlation for each $d\ts{A}$.

Figure~\ref{fig:MIOmegadhaplots}(c) displays the dependence of mutual information on $d\ts{A}$.
From Eqs.~\eqref{eq:local temperature} and \eqref{eq:redshift and distance}, the change in mutual information in this figure is caused by both particle production and redshift effects. 
Overall, as the detectors together move away from the black hole at constant proper separation, the mutual information initially grows very quickly, reaches a maximum, and then flattens to a constant corresponding to detectors in AdS$_3$ spacetime.

Recall  that a black hole inhibits static detectors from harvesting entanglement when one of them is close to the event horizon due to extreme Hawking radiation and redshift~\cite{henderson2018harvesting}. 
This region near the event horizon is called the ``entanglement shadow.'' 
In contrast to this we see in 
Fig.~\ref{fig:MIOmegadhaplots}(c) that mutual information is everywhere nonzero except at $d\ts{A}/\sigma=0$. 
Thus harvested correlation in the entanglement shadow contains either classical correlation or nondistillable entanglement. 
These results are commensurate with previous studies in $(1+1)$ dimensions~\cite{Tjoa2020vaidya, Ken.Freefall.PhysRevD.104.025001}, in which mutual information had no shadow region for all $\Omega \sigma$.

As suggested in Ref.~\cite{henderson2018harvesting}, we expect that the decline of correlation near the event horizon and its death at $d\ts{A}/\sigma=0$ are caused by the gravitational redshift and extreme Hawking radiation. 
From now on, we examine how these two effects contribute by treating detector A's redshift factor $\gamma\ts{A}$ and KMS temperature $T\ts{A}$ as independent variables, and write mutual information as a function of them: $I\ts{AB}=I\ts{AB}(T\ts{A}, \gamma\ts{A})$. 
We will show that the decline of mutual information is caused by both high temperature and large redshift, but the death is purely due to an extreme temperature effect.

\begin{figure*}[tp]
    \centering
    \includegraphics[width=\linewidth]{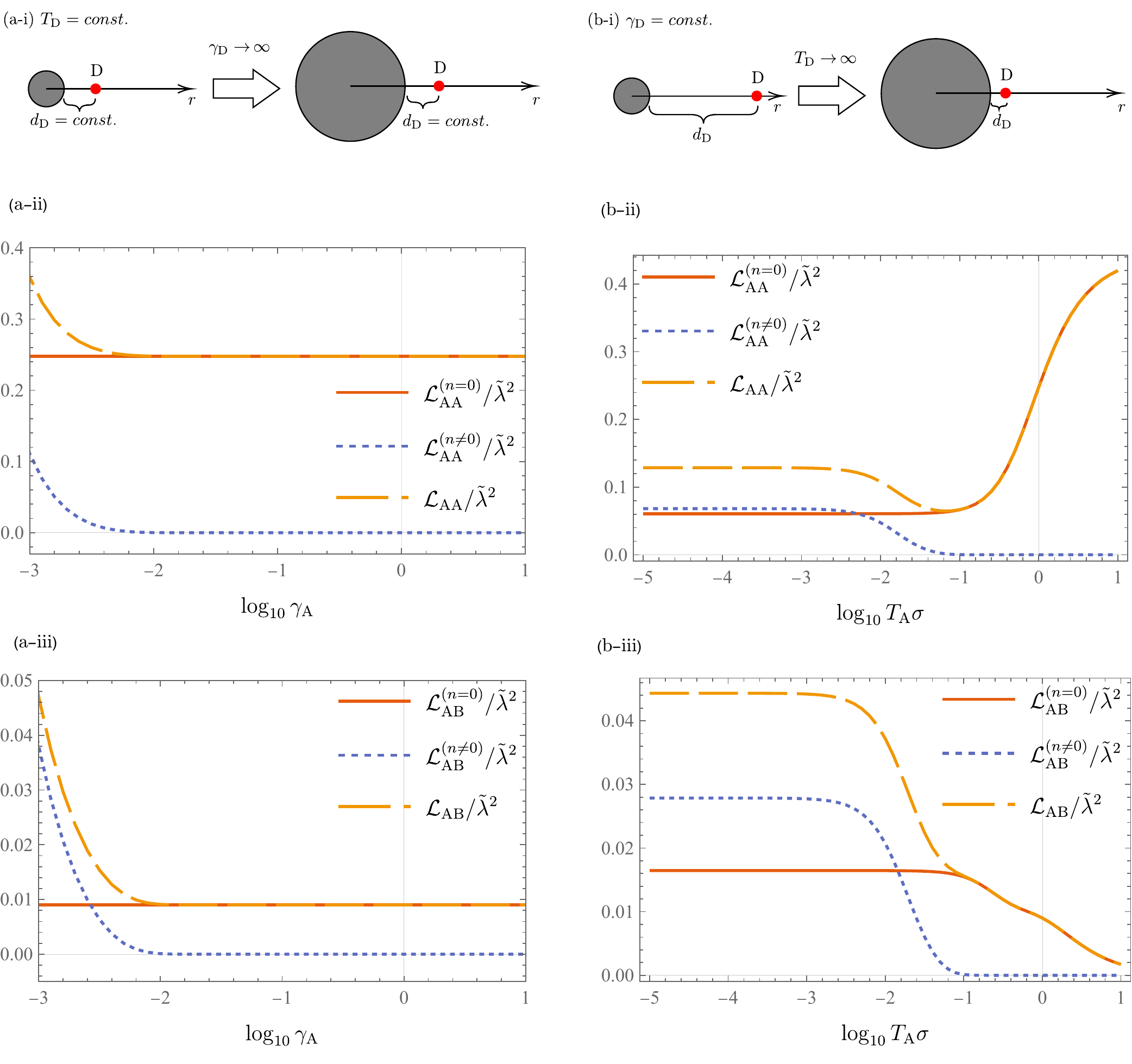}\\
    \caption{
    ~AdS-Rindler and BTZ contributions in $\mathcal{L}_{ij}$. 
    All figures have $\ell/\sigma=10, d\ts{A}/\sigma=7$, and $\Omega\sigma=1$. 
    (a-i)-(a-iii): Varying $\gamma\ts{A}$ while $T\ts{A}\sigma=1$ is fixed. 
    For both $\mathcal{L}\ts{AA}$ and $\mathcal{L}\ts{AB}$, the AdS-Rindler part ($n=0$) is independent of $\gamma\ts{A}$ and the BTZ part ($n\neq 0$) is nonzero only for $\gamma\ts{A}\ll 1$. 
    Because of the AdS-Rindler contribution, the total $\mathcal{L}_{ij}$ is nonzero for all $\gamma\ts{A}$. 
    (b-i)-(b-iii): Varying $T\ts{A}$ with fixed $\gamma\ts{A}=1/10$. 
    For $\mathcal{L}\ts{AB}$, both AdS-Rindler and BTZ terms vanish at high temperature. }
    \label{fig:AdSRindlervsBTZ}
\end{figure*}

Our strategy is to use $I\ts{AB}=I\ts{AB}(T\ts{A}, \gamma\ts{A})$ and fix either $T\ts{A}$ or $\gamma\ts{A}$ so that one can see respectively the influence of $\gamma\ts{A}$ and $T\ts{A}$. 
To this end, one needs to write the event horizon $r\ts{h}$ and the proper distance $d\ts{A}$ in terms of $T\ts{A}$ and $\gamma\ts{A}$: 
\begin{align}
    r\ts{h}
    &=
        2\pi \ell^2 T\ts{A} \gamma\ts{A}\,, \\
    d\ts{A}
    &= 
        \ell \ln 
        \dfrac{ 1 + \sqrt{ 1+ (2\pi \ell T\ts{A})^2 } }{ 2\pi \ell T\ts{A} }\,.
\end{align}
Thus the scenario where $T\ts{A}=const$ is different from that of $\gamma\ts{A}=const$ as illustrated in Fig.~\ref{fig:AdSRindlervsBTZ}. 
Fixing $T\ts{A}$ and varying $\gamma\ts{A}$ corresponds to a situation in Fig.~\ref{fig:AdSRindlervsBTZ}(a-i) where the size of the  black hole varies with $\gamma\ts{A}$ and the proper distance $d\ts{A}$ does not change. 
Note that the local temperature detected by detector B, $T\ts{B}$, is also fixed. 
Meanwhile, setting $\gamma\ts{A}$ a constant and varying $T\ts{A}$ means both $r\ts{h}$ and $d\ts{A}$ change with $T\ts{A}$ as shown in Fig.~\ref{fig:AdSRindlervsBTZ}(b-i).

\begin{figure*}[tp]
    \centering
    \includegraphics[width=\linewidth]{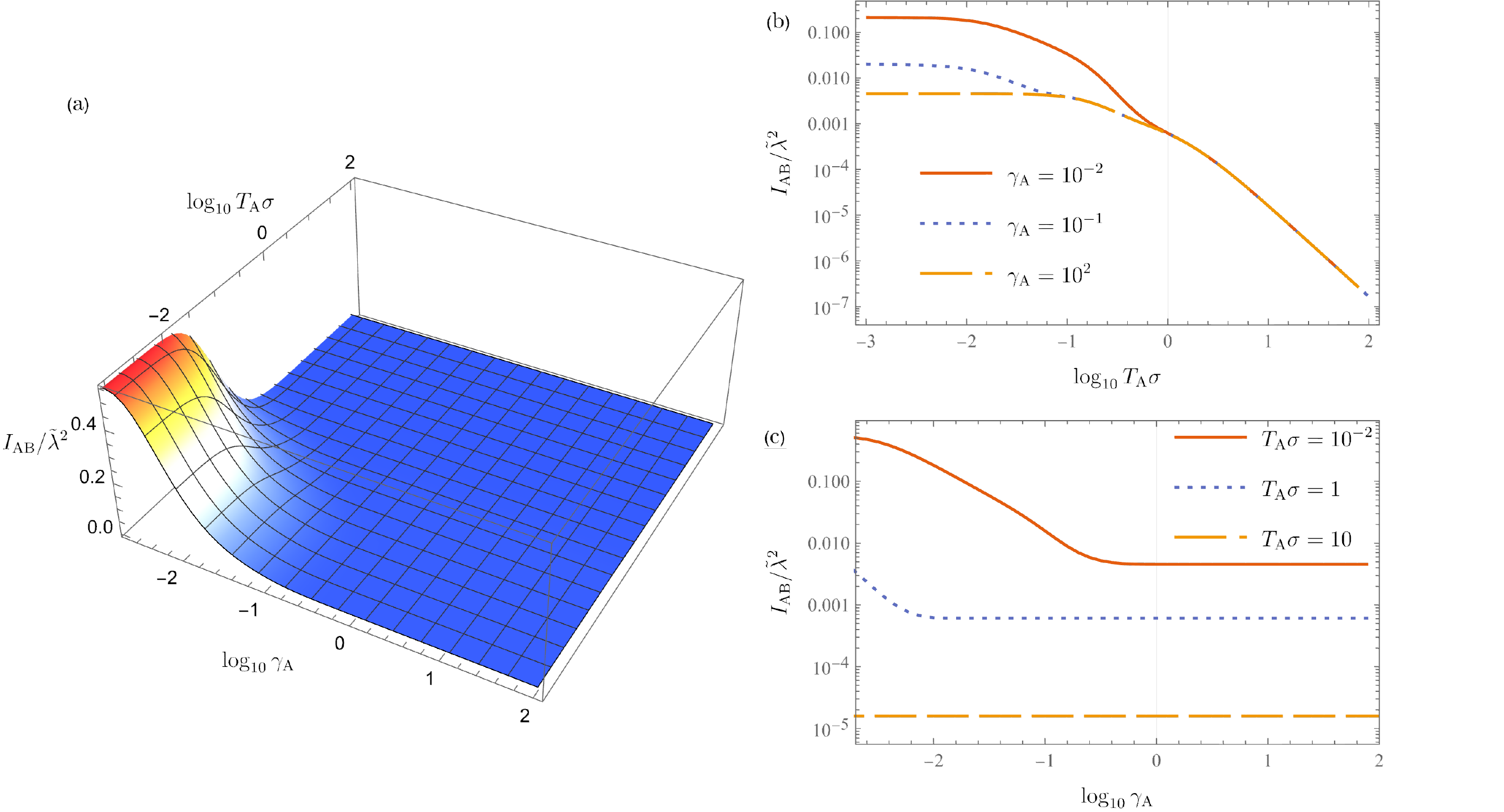}\\
    \caption{
    ~(a) 3D plot of mutual information $I\ts{AB}/\tilde{\lambda}^2$ as a function of local temperature $T\ts{A}$ and redshift factor $\gamma\ts{A}$ of detector A. 
    Here, $\ell/\sigma=10, \Omega\sigma=1$, and $d\ts{AB}/\sigma=7$. 
    (b) Slices of (a) at constant $\gamma\ts{A}$ in a logarithmic scale. 
    The curves in the lower temperature show the scenario where static detectors in AdS$_3$ spacetime without radiation, and the higher temperature regime represents how Unruh and/or Hawking radiations ``attack'' the detectors. 
    (c) Slices of (a) at constant $T\ts{A}\sigma$ in a logarithmic scale. 
    The contribution coming from the black hole can be seen in $\gamma\ts{A}\ll 1$ regime. 
    As $\gamma\ts{A}\to \infty$, the Hawking effect becomes negligible and only Unruh effect survives. }
    \label{fig:MItempRedshift}
\end{figure*}

Using the setup given above, let us take a look at the AdS-Rindler ($n=0$) and BTZ ($n\neq 0$) contributions in the response function $\mathcal{L}\ts{AA}/\tilde{\lambda}^2$ and the off-diagonal element $\mathcal{L}\ts{AB}/\tilde{\lambda}^2$. 
Figures~\ref{fig:AdSRindlervsBTZ}(a-ii) and (a-iii) are $\mathcal{L}\ts{AA}/\tilde{\lambda}^2$ and $\mathcal{L}\ts{AB}/\tilde{\lambda}^2$ when the local KMS temperature of detector-A is fixed. 
One can observe that the AdS-Rindler term in $\mathcal{L}\ts{AA}$ and $\mathcal{L}\ts{AB}$ is independent of $\gamma\ts{A}$, whereas the BTZ part is nonvanishing only when $\gamma\ts{A}\ll 1$. 
Hence, $\mathcal{L}_{ij}=\mathcal{L}_{ij}^{(n=0)} + \mathcal{L}_{ij}^{(n\neq 0)}$ is nonzero for all $\gamma\ts{A}$, indicating that mutual information will not vanish because of the gravitational redshift. 

On the other hand, Figs.~\ref{fig:AdSRindlervsBTZ}(b) depict the matrix elements $\mathcal{L}_{ij}$ when the redshift factor $\gamma\ts{A}$ is fixed. 
We first note that the response function $\mathcal{L}\ts{AA}$ of detector A [Fig.~\ref{fig:AdSRindlervsBTZ}(b-ii)] 
decreases as the temperature increases.
In a nutshell, one expects the contrary: that a detector's response function increases monotonically as the local temperature increases. 
However, this turns out to be not necessarily true; even a carefully switched detector with an infinite interaction duration experiences ``cool down'' as temperature increases. More precisely,
writing $\mathcal{F}(\Omega) = \mathcal{L}\ts{DD}/\tilde{\lambda}^2$, 
 the conditions
\begin{align}\label{weakAH}
&\frac{\dd\mathcal{F}(\Omega)}{\dd T\ts{KMS}}<0  \quad \textrm{weak}\\
&  \frac{\partial T\ts{EDR}}{\partial T\ts{KMS}} < 0
\quad  
\textrm{strong} 
\label{strongAH}
\end{align} 
in the presence of a black hole are respectively referred to as the weak and strong anti-Hawking effects \cite{Henderson2019anti-hawking}
(see also \cite{Campos-hyperbolicBH, Campos-RobinBC, Robbins-Anti-Hawking, Conroy-extremealBH}), where
\begin{align}
T\ts{EDR}=-\frac{\Omega}{\ln\mathcal{R}}
\end{align}
with
 \begin{align}
 \mathcal{R}=\frac{\mathcal{F}(\Omega)}{\mathcal{F}(-\Omega)}\,,
 \end{align}
 being the excitation-to-de-excitation ratio of the detector. For an accelerating detector in a flat spacetime these effects are respectively referred to as weak and strong anti-Unruh phenomena
\cite{Brenna2016anti-unruh, Garay2016anti-unruh}, and   will not be present in a situation where an inertial detector is in a thermal bath.

Since there is a range in temperature in which $\mathcal{L}\ts{AA}/\tilde{\lambda}^2$ decreases, the detector exhibits the weak anti-Hawking effect and the redshift is not involved. 
Also note that (under the Dirichlet boundary condition) this originates from the BTZ part of the Wightman function, as demonstrated in \cite{Henderson2019anti-hawking}. 

Figure~\ref{fig:AdSRindlervsBTZ}(b-iii) shows that the off-diagonal term $\mathcal{L}\ts{AB}/\tilde{\lambda}^2$ asymptotes to 0 as $T\ts{A}\sigma\to \infty$. 
This suggests that it is extreme temperature that inhibits detectors from harvesting mutual information. 
Hence, the fact that the detectors cannot extract correlations when one of them is at the event horizon [Figs.~\ref{fig:MIOmegadhaplots}(a), (c)] is purely due to the extremity of the local Hawking radiation there, which is a combination of increasing black hole mass with decreasing $d\ts{A}$ so as to ensure constant redshift.

Finally, we plot mutual information $I\ts{AB}$ as a function of detector A's local temperature, $T\ts{A}$, and the redshift factor, $\gamma\ts{A}$, in Fig.~\ref{fig:MItempRedshift}(a). 
Here we fix $\ell/\sigma=10$, $d\ts{AB}/\sigma=7$, and $\Omega \sigma=1$. 
One can observe from Fig. \ref{fig:MItempRedshift}(a) that the detectors easily harvest mutual information when the temperature $T\ts{A}$ and redshift $\gamma\ts{A}$ are both small. 
This corresponds to a case where two detectors are located far away ($d\ts{A}/\sigma \gg 1$) from a tiny black hole ($r\ts{h}/\sigma \ll 1$). 
On the other hand, high temperature or large redshift factor, which corresponds to $r\ts{h}/\sigma \gg 1$ and $d\ts{A}/\sigma \ll 1$, suppress the extraction of correlation. 

Figure~\ref{fig:MItempRedshift}(b) depicts slices of Fig.~\ref{fig:MItempRedshift}(a) with constant $\gamma\ts{A}$ in a logarithmic scale to analyze the relationship of mutual information with the Hawking effect. 
As the figure indicates, the colder the temperature, the more the correlation harvested, whereas $I\ts{AB}/\tilde{\lambda}^2 \to 0$ as the temperature gets large. 
This indicates that the radiation from the black hole acts as noise that inhibits  extraction of  correlation by the detectors,  and this is true no matter what the value of redshift is.

Conversely, Fig.~\ref{fig:MItempRedshift}(c) exhibits the influence of redshift $\gamma\ts{A}$ while the temperature $T\ts{A}$ is fixed. 
In contrast to Fig.~\ref{fig:MItempRedshift}(b), mutual information does not go to 0 as $\gamma\ts{A}\to \infty$. 
Instead, it asymptotes to a finite value that is characterized by the  temperature $T\ts{A}$. 
As we saw in Fig.~\ref{fig:AdSRindlervsBTZ}(a-ii) and (a-iii), the BTZ part of $\mathcal{L}_{ij}$ vanishes at large $\gamma\ts{A}$, leading us to conclude that the asymptotic values in Fig.~\ref{fig:MItempRedshift}(c) coincide with mutual information harvested by \textit{accelerating} detectors in AdS$_3$ spacetime with corresponding accelerations that give local temperature $T\ts{A}$. 
From these figures, we conclude that the death of mutual information in a black hole spacetime is purely due to Hawking radiation. 
We have checked that this feature holds for other boundary conditions ($\zeta=0,-1$).

\section{Conclusion}
\label{sec: conclusion}
Based on the fact that a vacuum state of a quantum field is entangled and that such correlations contain information about the background spacetime, correlation harvesting protocols are of great interest in relativistic quantum information theory. 
In this paper, we investigated the harvesting protocol for mutual information with two static UDW detectors in a nonrotating BTZ black hole spacetime. 

As shown in Refs.~\cite{henderson2018harvesting, Tjoa2020vaidya, Ken.Freefall.PhysRevD.104.025001, robbins2020entanglement}, there exists a region, the so-called entanglement shadow, near a black hole where static detectors cannot extract entanglement from the vacuum. 
It was suggested that the entanglement shadow is a consequence of the Hawking effect and gravitational redshift \cite{henderson2018harvesting}. 
Inspired by these results, we have examined the effect of Hawking radiation and gravitational redshift on harvested mutual information $I\ts{AB}$ between detectors A and B by considering $I\ts{AB}$ as a function of local temperature $T\ts{A}$ and redshift factor $\gamma\ts{A}$ of detector A: $I\ts{AB}=I\ts{AB}(T\ts{A}, \gamma\ts{A})$. 
 We find that the black hole has a significant effect on the mutual information extracted, with high temperature and extreme redshift preventing the detectors from extracting correlation.

We first confirmed that, unlike entanglement, there is no ``mutual information shadow'' in agreement with Refs. \cite{Tjoa2020vaidya, Ken.Freefall.PhysRevD.104.025001}. 
Mutual information vanishes only when one of the detectors reaches the event horizon, and this is true for any value of energy gap $\Omega$. 
We then showed that, by looking at the effects of redshift and Hawking radiation separately, extreme Hawking radiation 
(and not  gravitational redshift)  
is responsible for the death of mutual information . 
That is, as the local temperature increases while fixing $\gamma\ts{A}$, we found $I\ts{AB}\to 0$. 

Remarkably, the death of mutual information at extreme temperature is in contrast to what was found in flat spacetime \cite{simidzija2018harvesting}.  
Two inertial detectors interacting with a thermal field state in Minkowski spacetime struggle to extract entanglement but easily harvest mutual information as the field temperature increases. 
Since this result   holds for any spacetime dimension, switching functions, and detector shapes \cite{simidzija2018harvesting}, the striking distinction between our result from theirs may come from  (i) the existence of curvature, and/or (ii) the effects of acceleration. 
Concerning (i), the background spacetime has no curvature in \cite{simidzija2018harvesting} whereas the BTZ spacetime has  constant negative curvature. 
One can check the relevance of curvature  by putting inertial detectors in AdS$_3$ spacetime, which we do not investigate in this paper. 
Concerning (ii),   our static detectors in BTZ spacetime are constantly accelerating away from a black hole, whereas the previous study considered
 inertial detectors \cite{simidzija2018harvesting}.
A single detector can exhibit the anti-Unruh effect when accelerating, but not if it is inertial in a thermal bath \cite{Garay2016anti-unruh}. 
In the context of entanglement harvesting, detectors accelerating in a vacuum state and inertial in a thermal state can be distinguished \cite{salton2015acceleration,Liu:2021dnl}. 
It would be interesting to compare mutual information in scenarios with and without acceleration in the same spacetime.

\section*{Acknowledgments}

This work was supported in part by the Natural Sciences and Engineering Research Council of Canada and by Asian Office of Aerospace Research and Development Grant No. FA2386-19-1-4077.

\begin{widetext}

\appendix
\section{DERIVATION OF $\mathcal{L}\ts{AB}$}\label{app:Derivation of Lij}
Consider static detectors on the same axis originating at the center of the black hole
(i.e., $\Delta \phi=0$). 
Assuming that both detectors switch at the same time, we have 
\begin{align}
    \mathcal{L}\ts{AB}
    &=
        \lambda^2
        \int_{\mathbb{R}} \dd \tau\ts{A} 
        \int_{\mathbb{R}} \dd \tau\ts{B}\,
        e^{ -\tau\ts{A}^2/2\sigma^2 } e^{ -\tau\ts{B}^2/2\sigma^2 }
        e^{ -\ii \Omega (\tau\ts{A} - \tau\ts{B}) } 
        W\ts{BTZ}(\sx\ts{A}( \tau\ts{A}), \sx\ts{B}(\tau\ts{B} )  ),
\end{align}
where
\begin{align}
    &W\ts{BTZ}(\sx\ts{A}(\tau\ts{A}), \sx\ts{B}(\tau\ts{B}))
    =
        \dfrac{1}{ 4\pi \sqrt{2} \ell }
        \sum_{n=-\infty}^\infty 
        \kagikako{
            \dfrac{1}{ \sqrt{ \sigma_\epsilon (\sx\ts{A}, \Gamma^n \sx\ts{B}) } }
            -
            \dfrac{\zeta}{ \sqrt{ \sigma_\epsilon (\sx\ts{A}, \Gamma^n \sx\ts{B})+2 } }
        }, \\
    &\sigma_\epsilon (\sx\ts{A}, \Gamma^n \sx\ts{B})
    = 
        \dfrac{r\ts{A} r\ts{B}}{ r\ts{h}^2} 
        \cosh 
        \kako{
            \dfrac{r\ts{h}}{\ell} 2\pi n
        }
        -1 
        -\dfrac{ \sqrt{ (r\ts{A}^2-r\ts{h}^2) (r\ts{B}^2-r\ts{h}^2) } }{ r\ts{h}^2 }
        \cosh 
        \kagikako{
            \dfrac{r\ts{h}}{\ell^2} 
            \kako{ 
                \dfrac{\tau\ts{A}}{\gamma\ts{A}} - \dfrac{\tau\ts{B}}{\gamma\ts{B}}
            } 
            -\ii \epsilon
        },
\end{align}
where we used $t_j=\tau_j/\gamma_j$. 
Then $\mathcal{L}\ts{AB}$ becomes
\begin{align}
    \mathcal{L}\ts{AB}
    &=
        \dfrac{ \lambda^2 }{ 4\pi \sqrt{2} \ell } 
        \sum_{n=-\infty}^\infty 
        \int_{\mathbb{R}} \dd \tau\ts{A} 
        \int_{\mathbb{R}} \dd \tau\ts{B}\,
        e^{ -\tau\ts{A}^2/2\sigma^2 } e^{ -\tau\ts{B}^2/2\sigma^2 }
        e^{ -\ii \Omega (\tau\ts{A} - \tau\ts{B} ) } 
        \kagikako{
            \dfrac{1}{ \rho^-(\sx\ts{A}, \Gamma^n \sx\ts{B})} - \dfrac{\zeta}{ \rho^+(\sx\ts{A}, \Gamma^n \sx\ts{B}) } 
        } , 
\end{align}
where 
\begin{align}
    \rho^\pm (\sx\ts{A}, \Gamma^n \sx\ts{B})
    &\coloneqq
        \sqrt{
            \dfrac{r\ts{A} r\ts{B}}{ r\ts{h}^2} 
            \cosh 
            \kagikako{
                \dfrac{r\ts{h}}{\ell} 2\pi n
            }
            \pm 1 
            -\dfrac{ \ell^2 \gamma\ts{A} \gamma\ts{B} }{ r\ts{h}^2 }
            \cosh 
            \kagikako{
                \dfrac{r\ts{h}}{\ell^2} 
                \kako{
                    \dfrac{ \tau\ts{A} }{\gamma\ts{A}} 
                    - 
                    \dfrac{ \tau\ts{B} }{\gamma\ts{B}}
                }
                -\ii \epsilon
            }
        } \\
    &=
        \dfrac{\ell \sqrt{ \gamma\ts{A} \gamma\ts{B} }}{ r\ts{h} }
        \sqrt{
            \cosh \alpha^\pm\ts{AB$,n$}
            -\cosh 
            \kagikako{
                \dfrac{r\ts{h}}{\ell^2} 
                \kako{
                    \dfrac{ \tau\ts{A} }{\gamma\ts{A}} 
                    - 
                    \dfrac{ \tau\ts{B} }{\gamma\ts{B}}
                }
                -\ii \epsilon
            }
        } ,
\end{align}
and 
\begin{align}
    \alpha^\pm\ts{AB$,n$}
    &\coloneqq 
        \text{arccosh}
        \kagikako{
            \dfrac{r\ts{h}^2}{ \ell^2 \gamma\ts{A} \gamma\ts{B} }
            \kako{
                \dfrac{r\ts{A} r\ts{B}}{ r\ts{h}^2 }
                \cosh 
                \kagikako{
                    \dfrac{r\ts{h}}{\ell} 2\pi n
                }
                \pm 1
            }
        }. 
\end{align}
Let us change the variables $\tau\ts{A}$ and $\tau\ts{B}$ to $t\ts{A}$ and $t\ts{B}$, respectively, by the transformation $t_j \coloneqq \tau_j/\gamma_j$. 
\begin{align}
    \mathcal{L}\ts{AB}
    &=
        \dfrac{ \lambda^2 \gamma\ts{A} \gamma\ts{B} }{ 4\pi \sqrt{2} \ell } 
        \sum_{n=-\infty}^\infty 
        \int_{\mathbb{R}} \dd t\ts{A} \dd t\ts{B}\,
        e^{ -\gamma\ts{A}^2 t\ts{A}^2/2\sigma^2 } 
        e^{ -\gamma\ts{B}^2 t\ts{B}^2/2\sigma^2 }
        e^{ -\ii \Omega (\gamma\ts{A} t\ts{A} - \gamma\ts{B} t\ts{B} ) } 
        \kagikako{
            \dfrac{1}{ \rho^-(t\ts{A}, t\ts{B})} 
            - 
            \dfrac{\zeta}{ \rho^+(t\ts{A}, t\ts{B}) } 
        },
\end{align}
where 
\begin{align}
    \rho^\pm(t\ts{A}, t\ts{B})
    &\coloneqq
        \dfrac{\ell \sqrt{ \gamma\ts{A} \gamma\ts{B} }}{ r\ts{h} }
        \sqrt{
            \cosh \alpha^\pm\ts{AB$,n$} 
            -\cosh 
            \kagikako{
                \dfrac{r\ts{h}}{\ell^2} 
                \kako{
                    t\ts{A} - t\ts{B}
                }
                -\ii \epsilon
            }
        } . 
\end{align}
Further  changing  coordinates   $u\coloneqq t\ts{A}-t\ts{B}, s\coloneqq t\ts{A} + t\ts{B}$, the   corresponding Jacobian determinant is $J=1/2$, and so $\mathcal{L}\ts{AB}$ becomes 
\begin{align}
    \mathcal{L}\ts{AB}
    &=
        \dfrac{ \lambda^2 \gamma\ts{A} \gamma\ts{B} }{ 8\pi \sqrt{2} \ell } 
        \sum_{n=-\infty}^\infty 
        \int_{\mathbb{R}} \dd u\,
        e^{ -(\gamma\ts{A}^2 + \gamma\ts{B}^2)u^2/8\sigma^2 }
        e^{ -\ii \Omega (\gamma\ts{A} + \gamma\ts{B})u/2 }
        \kagikako{
            \dfrac{1}{ \rho^-(u)} 
            - 
            \dfrac{\zeta}{ \rho^+(u) } 
        } \notag \\
        &\times 
        \int_{\mathbb{R}} \dd s\,
        e^{ -(\gamma\ts{A}^2 + \gamma\ts{B}^2)s^2/8\sigma^2 }
        e^{ -(\gamma\ts{A}^2 - \gamma\ts{B}^2) us / 4\sigma^2 }
        e^{ -\ii \Omega (\gamma\ts{A} - \gamma\ts{B})s/2 } \\
    &=
        \dfrac{ \lambda^2 \sigma \gamma\ts{A} \gamma\ts{B} }{ 4 \ell \sqrt{\pi} \sqrt{ \gamma\ts{A}^2 + \gamma\ts{B}^2 } }
        \exp
        \kagikako{
            -\dfrac{ \Omega^2 \sigma^2 (\gamma\ts{A}-\gamma\ts{B})^2 }{ 2 (\gamma\ts{A}^2 + \gamma\ts{B}^2) }
        } 
        \sum_{n=-\infty}^\infty 
        \int_{\mathbb{R}} \dd u\,
        e^{ -a u^2 } 
        e^{ -\ii \beta u }
        \kagikako{
            \dfrac{1}{ \rho^-(u)} 
            - 
            \dfrac{\zeta}{ \rho^+(u) } 
        } \\
    &=
        2 K \sum_{n=-\infty}^\infty 
        \text{Re}\int_0^\infty \dd x\,
        e^{ -a x^2 }
        e^{ -\ii \beta x } 
        \kagikako{
            \dfrac{ 1 }{ \sqrt{ \cosh \alpha\ts{AB$,n$}^- - \cosh x } }
            -
            \dfrac{ \zeta }{ \sqrt{ \cosh \alpha\ts{AB$,n$}^+ - \cosh x } }
        }, \label{eq:LabAppendix}
\end{align}
where 
\begin{subequations}
\begin{align}
    a
    &\coloneqq
        \dfrac{ \gamma\ts{A}^2 \gamma\ts{B}^2 }{ 2\sigma^2 (\gamma\ts{A}^2 + \gamma\ts{B}^2) } 
        \dfrac{ \ell^4 }{ r\ts{h}^2 }, \\
    \beta
    &\coloneqq
        \dfrac{ \gamma\ts{A} \gamma\ts{B} (\gamma\ts{A} + \gamma\ts{B}) }{ \gamma\ts{A}^2 + \gamma\ts{B}^2 }
        \dfrac{ \ell^2 }{r\ts{h}}
        \Omega\,, \\
    K
    &\coloneqq
        \dfrac{ \lambda^2 \sigma \sqrt{ \gamma\ts{A} \gamma\ts{B} } }{ 4 \sqrt{\pi} \sqrt{ \gamma\ts{A}^2 + \gamma\ts{B}^2 } }
        \exp
        \kagikako{
            -\dfrac{ \Omega^2 \sigma^2 (\gamma\ts{A}-\gamma\ts{B})^2 }{ 2 (\gamma\ts{A}^2 + \gamma\ts{B}^2) }
        } .
\end{align}

\end{subequations}
Although the integrand in Eq.~\eqref{eq:LabAppendix} does not have a pole, one needs to be careful about the branch cuts as shown in Fig.~\ref{fig:contour}. 
Equation \eqref{eq:LabAppendix} can be obtained by evaluating along $L_1, C_\rho$, and $L_2$ in the limit $R\to \infty$ and $\rho\to 0$, where $\rho$ is the radius of the semicircle around the branch point. 

Let us choose a contour in the complex plane and use the Cauchy integral theorem. 
To do so, consider a complex integral of Eq. \eqref{eq:LabAppendix}:
\begin{align}
        \oint_C \dd z\,
        e^{ -a z^2 } e^{ -\ii \beta z } 
        \kagikako{
            \dfrac{ 1 }{ \sqrt{ \cosh \alpha\ts{AB$,n$}^- - \cosh z } }
            -
            \dfrac{ \zeta }{ \sqrt{ \cosh \alpha\ts{AB$,n$}^+ - \cosh z } }
        }\,.\label{eq:complex integral}
\end{align}
Assuming $\Omega>0$, the integral \eqref{eq:complex integral} converges if $| e^{-az^2} e^{ -\ii \beta z } |=e^{ -a(x^2 -y^2) } e^{ \beta y }<1$, where we used $z=x+\ii y$, and so the contour $C$ should be in $y<0$ and $-x \leq y \leq x$. 
From this, we choose the contour $C=L_1 C_\rho L_2 L_3 L_4$ as shown in Fig.~\ref{fig:contour} and by making use of the Cauchy integral theorem with the fact that $\lim_{R\to \infty} \int_{L_3}= 0$, Eq. \eqref{eq:LabAppendix} becomes an integral along $L_4^{-1}$: 
\begin{align}
    \mathcal{L}\ts{AB}
    &=
        2 K \sum_{n=-\infty}^\infty 
            \lim_{R\to \infty}
            \text{Re}
            \int_0^{R+\ii \eta} \dd z\,
            e^{ -a z^2 }
            e^{ -\ii \beta z } 
            \kagikako{
                \dfrac{ 1 }{ \sqrt{ \cosh \alpha\ts{AB$,n$}^- - \cosh z } }
                -
                \dfrac{ \zeta }{ \sqrt{ \cosh \alpha\ts{AB$,n$}^+ - \cosh z } }
            }, 
\end{align}
where $\eta \in (-\pi,0)$ and $\Omega>0$; this is the
expression we compute numerically. 

\begin{figure}[t]
    \centering
    \includegraphics[width=7cm]{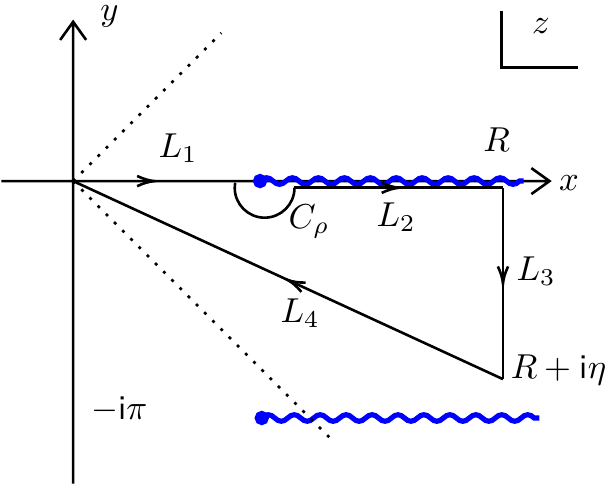}\\
    \caption{
    ~Contour in a complex plane. 
    Blue wiggly lines are the branch cuts and the contour is chosen so that the integral converges when $\Omega>0$. 
    The dotted lines represent $y=\pm x$, and the contour must be inside of this region. }
    \label{fig:contour}
\end{figure}

For $\mathcal{L}\ts{AA}$ and $\mathcal{L}\ts{BB}$, they are known to be \cite{henderson2018harvesting}
\begin{align}
    \mathcal{L}\ts{DD}
    &=
        \dfrac{\lambda^2 \sigma^2}{ 2 }
        \int_{\mathbb{R}} \dd x\,
        \dfrac{ e^{ -\sigma^2 (x-\Omega)^2 } }{ e^{ x/T\ts{D} } +1 }
        -\zeta \dfrac{ \lambda^2 \sigma }{2 \sqrt{2 \pi} }
        \text{Re}
        \int_0^\infty \dd x\,
       \dfrac{ e^{-a\ts{D} x^2} e^{-\ii \beta\ts{D} x} }{ \sqrt{ \cosh \alpha\ts{D,0}^+ - \cosh x } }
       \notag \\
        &+\dfrac{ \lambda^2 \sigma }{ \sqrt{2 \pi} }
        \sum_{n=1}^\infty 
        \text{Re}
        \int_0^\infty \dd x\,
        e^{ -a\ts{D} x^2 } e^{ -\ii \beta\ts{D} x }
        \kako{
            \dfrac{1}{ \sqrt{ \cosh \alpha\ts{D$,n$}^- -\cosh x } }
            -
            \dfrac{\zeta}{ \sqrt{ \cosh \alpha\ts{D$,n$}^+ -\cosh x } }
        } , \label{eq:transition prob}
\end{align}
where $T\ts{D}=r\ts{h}/2 \pi \ell^2 \gamma\ts{D}$ is the local temperature at $r=r\ts{D}$ and 
\begin{align}
    a\ts{D}
    &\coloneqq 
        \dfrac{ \ell^4 \gamma\ts{D}^2 }{ 4\sigma^2 r\ts{h}^2 },~
        \beta\ts{D} \coloneqq 
        \dfrac{ \ell^2 \gamma\ts{D} \Omega }{ r\ts{h} }\,, \\
    \alpha\ts{D$,n$}^\pm
    &\coloneqq
        \text{arccosh}
        \kagikako{
            \dfrac{r\ts{h}^2}{ \gamma\ts{D}^2 \ell^2 }
            \kako{
                \dfrac{r^2}{ r\ts{h}^2 } \cosh 
                \kagikako{
                    \dfrac{r\ts{h}}{\ell} 2\pi n
                } \pm 1
            }
        } .
\end{align}
The first two terms, which correspond to $n=0$, are the so-called AdS-Rindler terms and the last term ($n\neq 0$) is known as the BTZ term. 
The second and third integrals in \eqref{eq:transition prob} also have the same branch cut subtlety, but can be treated in the same manner as for $\mathcal{L}\ts{AB}$. 
\vspace{-5mm}

\end{widetext}

\bibliography{ref}

\begin{thebibliography}{46}%
\makeatletter
\providecommand \@ifxundefined [1]{%
 \@ifx{#1\undefined}
}%
\providecommand \@ifnum [1]{%
 \ifnum #1\expandafter \@firstoftwo
 \else \expandafter \@secondoftwo
 \fi
}%
\providecommand \@ifx [1]{%
 \ifx #1\expandafter \@firstoftwo
 \else \expandafter \@secondoftwo
 \fi
}%
\providecommand \natexlab [1]{#1}%
\providecommand \enquote  [1]{``#1''}%
\providecommand \bibnamefont  [1]{#1}%
\providecommand \bibfnamefont [1]{#1}%
\providecommand \citenamefont [1]{#1}%
\providecommand \href@noop [0]{\@secondoftwo}%
\providecommand \href [0]{\begingroup \@sanitize@url \@href}%
\providecommand \@href[1]{\@@startlink{#1}\@@href}%
\providecommand \@@href[1]{\endgroup#1\@@endlink}%
\providecommand \@sanitize@url [0]{\catcode `\\12\catcode `\$12\catcode
  `\&12\catcode `\#12\catcode `\^12\catcode `\_12\catcode `\%12\relax}%
\providecommand \@@startlink[1]{}%
\providecommand \@@endlink[0]{}%
\providecommand \url  [0]{\begingroup\@sanitize@url \@url }%
\providecommand \@url [1]{\endgroup\@href {#1}{\urlprefix }}%
\providecommand \urlprefix  [0]{URL }%
\providecommand \Eprint [0]{\href }%
\providecommand \doibase [0]{https://doi.org/}%
\providecommand \selectlanguage [0]{\@gobble}%
\providecommand \bibinfo  [0]{\@secondoftwo}%
\providecommand \bibfield  [0]{\@secondoftwo}%
\providecommand \translation [1]{[#1]}%
\providecommand \BibitemOpen [0]{}%
\providecommand \bibitemStop [0]{}%
\providecommand \bibitemNoStop [0]{.\EOS\space}%
\providecommand \EOS [0]{\spacefactor3000\relax}%
\providecommand \BibitemShut  [1]{\csname bibitem#1\endcsname}%
\let\auto@bib@innerbib\@empty
\bibitem [{\citenamefont {Unruh}(1976)}]{Unruh1979evaporation}%
  \BibitemOpen
  \bibfield  {author} {\bibinfo {author} {\bibfnamefont {W.~G.}\ \bibnamefont
  {Unruh}},\ }\bibfield  {title} {\bibinfo {title} {Notes on black-hole
  evaporation},\ }\href {https://doi.org/10.1103/PhysRevD.14.870} {\bibfield
  {journal} {\bibinfo  {journal} {Phys. Rev. D}\ }\textbf {\bibinfo {volume}
  {14}},\ \bibinfo {pages} {870} (\bibinfo {year} {1976})}\BibitemShut
  {NoStop}%
\bibitem [{\citenamefont {{DeWitt}}(1979)}]{DeWitt1979}%
  \BibitemOpen
  \bibfield  {author} {\bibinfo {author} {\bibfnamefont {B.~S.}\ \bibnamefont
  {{DeWitt}}},\ }\bibfield  {title} {\bibinfo {title} {{Quantum gravity: The
  new synthesis}},\ }in\ \href@noop {} {\emph {\bibinfo {booktitle} {General
  Relativity: An Einstein Centenary Survey}}},\ \bibinfo {editor} {edited by\
  \bibinfo {editor} {\bibfnamefont {S.~W.}\ \bibnamefont {{Hawking}}}\ and\
  \bibinfo {editor} {\bibfnamefont {W.}~\bibnamefont {{Israel}}}}\ (\bibinfo
  {year} {1979})\ pp.\ \bibinfo {pages} {680--745}\BibitemShut {NoStop}%
\bibitem [{\citenamefont {Alsing}\ and\ \citenamefont
  {Milburn}(2003)}]{PhysRevLett.91.180404}%
  \BibitemOpen
  \bibfield  {author} {\bibinfo {author} {\bibfnamefont {P.~M.}\ \bibnamefont
  {Alsing}}\ and\ \bibinfo {author} {\bibfnamefont {G.~J.}\ \bibnamefont
  {Milburn}},\ }\bibfield  {title} {\bibinfo {title} {Teleportation with a
  uniformly accelerated partner},\ }\href
  {https://doi.org/10.1103/PhysRevLett.91.180404} {\bibfield  {journal}
  {\bibinfo  {journal} {Phys. Rev. Lett.}\ }\textbf {\bibinfo {volume} {91}},\
  \bibinfo {pages} {180404} (\bibinfo {year} {2003})}\BibitemShut {NoStop}%
\bibitem [{\citenamefont {Landulfo}\ and\ \citenamefont
  {Matsas}(2009)}]{Landulfo2009suddendeath}%
  \BibitemOpen
  \bibfield  {author} {\bibinfo {author} {\bibfnamefont {A.~G.~S.}\
  \bibnamefont {Landulfo}}\ and\ \bibinfo {author} {\bibfnamefont {G.~E.~A.}\
  \bibnamefont {Matsas}},\ }\bibfield  {title} {\bibinfo {title} {{Sudden death
  of entanglement and teleportation fidelity loss via the Unruh effect}},\
  }\href {https://doi.org/10.1103/PhysRevA.80.032315} {\bibfield  {journal}
  {\bibinfo  {journal} {Phys. Rev. A}\ }\textbf {\bibinfo {volume} {80}},\
  \bibinfo {pages} {032315} (\bibinfo {year} {2009})}\BibitemShut {NoStop}%
\bibitem [{\citenamefont {Cliche}\ and\ \citenamefont
  {Kempf}(2010)}]{Cliche2010channel}%
  \BibitemOpen
  \bibfield  {author} {\bibinfo {author} {\bibfnamefont {M.}~\bibnamefont
  {Cliche}}\ and\ \bibinfo {author} {\bibfnamefont {A.}~\bibnamefont {Kempf}},\
  }\bibfield  {title} {\bibinfo {title} {Relativistic quantum channel of
  communication through field quanta},\ }\href
  {https://doi.org/10.1103/PhysRevA.81.012330} {\bibfield  {journal} {\bibinfo
  {journal} {Phys. Rev. A}\ }\textbf {\bibinfo {volume} {81}},\ \bibinfo
  {pages} {012330} (\bibinfo {year} {2010})}\BibitemShut {NoStop}%
\bibitem [{\citenamefont {Jonsson}(2017)}]{jonsson2017quantum}%
  \BibitemOpen
  \bibfield  {author} {\bibinfo {author} {\bibfnamefont {R.~H.}\ \bibnamefont
  {Jonsson}},\ }\bibfield  {title} {\bibinfo {title} {Quantum signaling in
  relativistic motion and across acceleration horizons},\ }\href
  {https://doi.org/10.1088/1751-8121/aa7d3c} {\bibfield  {journal} {\bibinfo
  {journal} {Journal of Physics A: Mathematical and Theoretical}\ }\textbf
  {\bibinfo {volume} {50}},\ \bibinfo {pages} {355401} (\bibinfo {year}
  {2017})}\BibitemShut {NoStop}%
\bibitem [{\citenamefont {Landulfo}(2016)}]{Landulfo2016magnus1}%
  \BibitemOpen
  \bibfield  {author} {\bibinfo {author} {\bibfnamefont {A.~G.~S.}\
  \bibnamefont {Landulfo}},\ }\bibfield  {title} {\bibinfo {title}
  {Nonperturbative approach to relativistic quantum communication channels},\
  }\href {https://doi.org/10.1103/PhysRevD.93.104019} {\bibfield  {journal}
  {\bibinfo  {journal} {Phys. Rev. D}\ }\textbf {\bibinfo {volume} {93}},\
  \bibinfo {pages} {104019} (\bibinfo {year} {2016})}\BibitemShut {NoStop}%
\bibitem [{\citenamefont {Simidzija}\ \emph {et~al.}(2020)\citenamefont
  {Simidzija}, \citenamefont {Ahmadzadegan}, \citenamefont {Kempf},\ and\
  \citenamefont {Mart\'{\i}n-Mart\'{\i}nez}}]{Simidzija2020transmit}%
  \BibitemOpen
  \bibfield  {author} {\bibinfo {author} {\bibfnamefont {P.}~\bibnamefont
  {Simidzija}}, \bibinfo {author} {\bibfnamefont {A.}~\bibnamefont
  {Ahmadzadegan}}, \bibinfo {author} {\bibfnamefont {A.}~\bibnamefont
  {Kempf}},\ and\ \bibinfo {author} {\bibfnamefont {E.}~\bibnamefont
  {Mart\'{\i}n-Mart\'{\i}nez}},\ }\bibfield  {title} {\bibinfo {title}
  {Transmission of quantum information through quantum fields},\ }\href
  {https://doi.org/10.1103/PhysRevD.101.036014} {\bibfield  {journal} {\bibinfo
   {journal} {Phys. Rev. D}\ }\textbf {\bibinfo {volume} {101}},\ \bibinfo
  {pages} {036014} (\bibinfo {year} {2020})}\BibitemShut {NoStop}%
\bibitem [{\citenamefont {Fuentes-Schuller}\ and\ \citenamefont
  {Mann}(2005)}]{FuentesAliceFalls}%
  \BibitemOpen
  \bibfield  {author} {\bibinfo {author} {\bibfnamefont {I.}~\bibnamefont
  {Fuentes-Schuller}}\ and\ \bibinfo {author} {\bibfnamefont {R.~B.}\
  \bibnamefont {Mann}},\ }\bibfield  {title} {\bibinfo {title} {Alice falls
  into a black hole: Entanglement in noninertial frames},\ }\href
  {https://doi.org/10.1103/PhysRevLett.95.120404} {\bibfield  {journal}
  {\bibinfo  {journal} {Phys. Rev. Lett.}\ }\textbf {\bibinfo {volume} {95}},\
  \bibinfo {pages} {120404} (\bibinfo {year} {2005})}\BibitemShut {NoStop}%
\bibitem [{\citenamefont {Alsing}\ \emph {et~al.}(2006)\citenamefont {Alsing},
  \citenamefont {Fuentes-Schuller}, \citenamefont {Mann},\ and\ \citenamefont
  {Tessier}}]{AlsingDiracFields}%
  \BibitemOpen
  \bibfield  {author} {\bibinfo {author} {\bibfnamefont {P.~M.}\ \bibnamefont
  {Alsing}}, \bibinfo {author} {\bibfnamefont {I.}~\bibnamefont
  {Fuentes-Schuller}}, \bibinfo {author} {\bibfnamefont {R.~B.}\ \bibnamefont
  {Mann}},\ and\ \bibinfo {author} {\bibfnamefont {T.~E.}\ \bibnamefont
  {Tessier}},\ }\bibfield  {title} {\bibinfo {title} {Entanglement of dirac
  fields in noninertial frames},\ }\href
  {https://doi.org/10.1103/PhysRevA.74.032326} {\bibfield  {journal} {\bibinfo
  {journal} {Phys. Rev. A}\ }\textbf {\bibinfo {volume} {74}},\ \bibinfo
  {pages} {032326} (\bibinfo {year} {2006})}\BibitemShut {NoStop}%
\bibitem [{\citenamefont {Valentini}(1991)}]{Valentini1991nonlocalcorr}%
  \BibitemOpen
  \bibfield  {author} {\bibinfo {author} {\bibfnamefont {A.}~\bibnamefont
  {Valentini}},\ }\bibfield  {title} {\bibinfo {title} {Non-local correlations
  in quantum electrodynamics},\ }\href
  {https://doi.org/https://doi.org/10.1016/0375-9601(91)90952-5} {\bibfield
  {journal} {\bibinfo  {journal} {Phys. Lett.}\ }\textbf {\bibinfo {volume}
  {153A}},\ \bibinfo {pages} {321 } (\bibinfo {year} {1991})}\BibitemShut
  {NoStop}%
\bibitem [{\citenamefont {Reznik}(2003)}]{reznik2003entanglement}%
  \BibitemOpen
  \bibfield  {author} {\bibinfo {author} {\bibfnamefont {B.}~\bibnamefont
  {Reznik}},\ }\bibfield  {title} {\bibinfo {title} {Entanglement from the
  vacuum},\ }\href {https://doi.org/https://doi.org/10.1023/A:1022875910744}
  {\bibfield  {journal} {\bibinfo  {journal} {Found. Phys.}\ }\textbf {\bibinfo
  {volume} {33}},\ \bibinfo {pages} {167} (\bibinfo {year} {2003})}\BibitemShut
  {NoStop}%
\bibitem [{\citenamefont {Reznik}\ \emph {et~al.}(2005)\citenamefont {Reznik},
  \citenamefont {Retzker},\ and\ \citenamefont {Silman}}]{reznik2005violating}%
  \BibitemOpen
  \bibfield  {author} {\bibinfo {author} {\bibfnamefont {B.}~\bibnamefont
  {Reznik}}, \bibinfo {author} {\bibfnamefont {A.}~\bibnamefont {Retzker}},\
  and\ \bibinfo {author} {\bibfnamefont {J.}~\bibnamefont {Silman}},\
  }\bibfield  {title} {\bibinfo {title} {{Violating Bell's inequalities in
  vacuum}},\ }\href {https://doi.org/10.1103/PhysRevA.71.042104} {\bibfield
  {journal} {\bibinfo  {journal} {Phys. Rev. A}\ }\textbf {\bibinfo {volume}
  {71}},\ \bibinfo {pages} {042104} (\bibinfo {year} {2005})}\BibitemShut
  {NoStop}%
\bibitem [{\citenamefont {Summers}\ and\ \citenamefont
  {Werner}(1985)}]{summers1985bell}%
  \BibitemOpen
  \bibfield  {author} {\bibinfo {author} {\bibfnamefont {S.~J.}\ \bibnamefont
  {Summers}}\ and\ \bibinfo {author} {\bibfnamefont {R.}~\bibnamefont
  {Werner}},\ }\bibfield  {title} {\bibinfo {title} {{The vacuum violates
  Bell's inequalities}},\ }\href
  {https://doi.org/https://doi.org/10.1016/0375-9601(85)90093-3} {\bibfield
  {journal} {\bibinfo  {journal} {Phys. Lett.}\ }\textbf {\bibinfo {volume}
  {110A}},\ \bibinfo {pages} {257 } (\bibinfo {year} {1985})}\BibitemShut
  {NoStop}%
\bibitem [{\citenamefont {Summers}\ and\ \citenamefont
  {Werner}(1987)}]{summers1987bell}%
  \BibitemOpen
  \bibfield  {author} {\bibinfo {author} {\bibfnamefont {S.~J.}\ \bibnamefont
  {Summers}}\ and\ \bibinfo {author} {\bibfnamefont {R.}~\bibnamefont
  {Werner}},\ }\bibfield  {title} {\bibinfo {title} {Bell’s inequalities and
  quantum field theory. i. general setting},\ }\href
  {https://doi.org/10.1063/1.527733} {\bibfield  {journal} {\bibinfo  {journal}
  {J. Math. Phys. (N.Y.)}\ }\textbf {\bibinfo {volume} {28}},\ \bibinfo {pages}
  {2440} (\bibinfo {year} {1987})}\BibitemShut {NoStop}%
\bibitem [{\citenamefont {Pozas-Kerstjens}\ and\ \citenamefont
  {Mart\'{\i}n-Mart\'{\i}nez}(2015)}]{pozas2015harvesting}%
  \BibitemOpen
  \bibfield  {author} {\bibinfo {author} {\bibfnamefont {A.}~\bibnamefont
  {Pozas-Kerstjens}}\ and\ \bibinfo {author} {\bibfnamefont {E.}~\bibnamefont
  {Mart\'{\i}n-Mart\'{\i}nez}},\ }\bibfield  {title} {\bibinfo {title}
  {Harvesting correlations from the quantum vacuum},\ }\href
  {https://doi.org/10.1103/PhysRevD.92.064042} {\bibfield  {journal} {\bibinfo
  {journal} {Phys. Rev. D}\ }\textbf {\bibinfo {volume} {92}},\ \bibinfo
  {pages} {064042} (\bibinfo {year} {2015})}\BibitemShut {NoStop}%
\bibitem [{\citenamefont {Mart\'{\i}n-Mart\'{\i}nez}\ \emph
  {et~al.}(2016)\citenamefont {Mart\'{\i}n-Mart\'{\i}nez}, \citenamefont
  {Smith},\ and\ \citenamefont {Terno}}]{smith2016topology}%
  \BibitemOpen
  \bibfield  {author} {\bibinfo {author} {\bibfnamefont {E.}~\bibnamefont
  {Mart\'{\i}n-Mart\'{\i}nez}}, \bibinfo {author} {\bibfnamefont {A.~R.~H.}\
  \bibnamefont {Smith}},\ and\ \bibinfo {author} {\bibfnamefont {D.~R.}\
  \bibnamefont {Terno}},\ }\bibfield  {title} {\bibinfo {title} {Spacetime
  structure and vacuum entanglement},\ }\href
  {https://doi.org/10.1103/PhysRevD.93.044001} {\bibfield  {journal} {\bibinfo
  {journal} {Phys. Rev. D}\ }\textbf {\bibinfo {volume} {93}},\ \bibinfo
  {pages} {044001} (\bibinfo {year} {2016})}\BibitemShut {NoStop}%
\bibitem [{\citenamefont {Kukita}\ and\ \citenamefont
  {Nambu}(2017)}]{kukita2017harvesting}%
  \BibitemOpen
  \bibfield  {author} {\bibinfo {author} {\bibfnamefont {S.}~\bibnamefont
  {Kukita}}\ and\ \bibinfo {author} {\bibfnamefont {Y.}~\bibnamefont {Nambu}},\
  }\bibfield  {title} {\bibinfo {title} {{Harvesting large scale entanglement
  in de Sitter space with multiple detectors}},\ }\href
  {https://doi.org/10.3390/e19090449} {\bibfield  {journal} {\bibinfo
  {journal} {Entropy}\ }\textbf {\bibinfo {volume} {19}},\ \bibinfo {pages}
  {449} (\bibinfo {year} {2017})}\BibitemShut {NoStop}%
\bibitem [{\citenamefont {Henderson}\ \emph {et~al.}(2018)\citenamefont
  {Henderson}, \citenamefont {Hennigar}, \citenamefont {Mann}, \citenamefont
  {Smith},\ and\ \citenamefont {Zhang}}]{henderson2018harvesting}%
  \BibitemOpen
  \bibfield  {author} {\bibinfo {author} {\bibfnamefont {L.~J.}\ \bibnamefont
  {Henderson}}, \bibinfo {author} {\bibfnamefont {R.~A.}\ \bibnamefont
  {Hennigar}}, \bibinfo {author} {\bibfnamefont {R.~B.}\ \bibnamefont {Mann}},
  \bibinfo {author} {\bibfnamefont {A.~R.~H.}\ \bibnamefont {Smith}},\ and\
  \bibinfo {author} {\bibfnamefont {J.}~\bibnamefont {Zhang}},\ }\bibfield
  {title} {\bibinfo {title} {Harvesting entanglement from the black hole
  vacuum},\ }\href {https://doi.org/10.1088/1361-6382/aae27e} {\bibfield
  {journal} {\bibinfo  {journal} {Classical Quantum Gravity}\ }\textbf
  {\bibinfo {volume} {35}},\ \bibinfo {pages} {21LT02} (\bibinfo {year}
  {2018})}\BibitemShut {NoStop}%
\bibitem [{\citenamefont {Ng}\ \emph {et~al.}(2018)\citenamefont {Ng},
  \citenamefont {Mann},\ and\ \citenamefont
  {Mart\'{\i}n-Mart\'{\i}nez}}]{ng2018AdS}%
  \BibitemOpen
  \bibfield  {author} {\bibinfo {author} {\bibfnamefont {K.~K.}\ \bibnamefont
  {Ng}}, \bibinfo {author} {\bibfnamefont {R.~B.}\ \bibnamefont {Mann}},\ and\
  \bibinfo {author} {\bibfnamefont {E.}~\bibnamefont
  {Mart\'{\i}n-Mart\'{\i}nez}},\ }\bibfield  {title} {\bibinfo {title}
  {{Unruh-DeWitt detectors and entanglement: The anti--de Sitter space}},\
  }\href {https://doi.org/10.1103/PhysRevD.98.125005} {\bibfield  {journal}
  {\bibinfo  {journal} {Phys. Rev. D}\ }\textbf {\bibinfo {volume} {98}},\
  \bibinfo {pages} {125005} (\bibinfo {year} {2018})}\BibitemShut {NoStop}%
\bibitem [{\citenamefont {Cong}\ \emph {et~al.}(2020)\citenamefont {Cong},
  \citenamefont {Qian}, \citenamefont {Good},\ and\ \citenamefont
  {Mann}}]{cong2020horizon}%
  \BibitemOpen
  \bibfield  {author} {\bibinfo {author} {\bibfnamefont {W.}~\bibnamefont
  {Cong}}, \bibinfo {author} {\bibfnamefont {C.}~\bibnamefont {Qian}}, \bibinfo
  {author} {\bibfnamefont {M.~R.}\ \bibnamefont {Good}},\ and\ \bibinfo
  {author} {\bibfnamefont {R.~B.}\ \bibnamefont {Mann}},\ }\bibfield  {title}
  {\bibinfo {title} {Effects of horizons on entanglement harvesting},\ }\href
  {https://doi.org/10.1007/JHEP10(2020)067} {\bibfield  {journal} {\bibinfo
  {journal} {J. High Energy Phys.}\ }\textbf {\bibinfo {volume} {10}}\bibinfo
  {number} { (2020)},\ \bibinfo {pages} {67}}\BibitemShut {NoStop}%
\bibitem [{\citenamefont {Gray}\ \emph {et~al.}(2021)\citenamefont {Gray},
  \citenamefont {Kubizňák}, \citenamefont {May}, \citenamefont {Timmerman},\
  and\ \citenamefont {Tjoa}}]{FinnShockwave}%
  \BibitemOpen
\bibfield  {number} {  }\bibfield  {author} {\bibinfo {author} {\bibfnamefont
  {F.}~\bibnamefont {Gray}}, \bibinfo {author} {\bibfnamefont {D.}~\bibnamefont
  {Kubizňák}}, \bibinfo {author} {\bibfnamefont {T.}~\bibnamefont {May}},
  \bibinfo {author} {\bibfnamefont {S.}~\bibnamefont {Timmerman}},\ and\
  \bibinfo {author} {\bibfnamefont {E.}~\bibnamefont {Tjoa}},\ }\bibfield
  {title} {\bibinfo {title} {{Quantum imprints of gravitational shockwaves}},\
  }\href {https://doi.org/https://doi.org/10.1007/JHEP11(2021)054} {\bibfield
  {journal} {\bibinfo  {journal} {J. High Energy Phys.}\ }\textbf {\bibinfo
  {volume} {11}}\bibinfo  {number} { (2021)},\ \bibinfo {pages}
  {054}}\BibitemShut {NoStop}%
\bibitem [{\citenamefont {Salton}\ \emph {et~al.}(2015)\citenamefont {Salton},
  \citenamefont {Mann},\ and\ \citenamefont
  {Menicucci}}]{salton2015acceleration}%
  \BibitemOpen
\bibfield  {number} {  }\bibfield  {author} {\bibinfo {author} {\bibfnamefont
  {G.}~\bibnamefont {Salton}}, \bibinfo {author} {\bibfnamefont {R.~B.}\
  \bibnamefont {Mann}},\ and\ \bibinfo {author} {\bibfnamefont {N.~C.}\
  \bibnamefont {Menicucci}},\ }\bibfield  {title} {\bibinfo {title}
  {Acceleration-assisted entanglement harvesting and rangefinding},\ }\href
  {https://doi.org/10.1088/1367-2630/17/3/035001} {\bibfield  {journal}
  {\bibinfo  {journal} {New J. Phys.}\ }\textbf {\bibinfo {volume} {17}},\
  \bibinfo {pages} {035001} (\bibinfo {year} {2015})}\BibitemShut {NoStop}%
\bibitem [{\citenamefont {Foo}\ \emph {et~al.}(2020)\citenamefont {Foo},
  \citenamefont {Onoe},\ and\ \citenamefont
  {Zych}}]{FooSuperpositionTrajectory}%
  \BibitemOpen
  \bibfield  {author} {\bibinfo {author} {\bibfnamefont {J.}~\bibnamefont
  {Foo}}, \bibinfo {author} {\bibfnamefont {S.}~\bibnamefont {Onoe}},\ and\
  \bibinfo {author} {\bibfnamefont {M.}~\bibnamefont {Zych}},\ }\bibfield
  {title} {\bibinfo {title} {Unruh-dewitt detectors in quantum superpositions
  of trajectories},\ }\href {https://doi.org/10.1103/PhysRevD.102.085013}
  {\bibfield  {journal} {\bibinfo  {journal} {Phys. Rev. D}\ }\textbf {\bibinfo
  {volume} {102}},\ \bibinfo {pages} {085013} (\bibinfo {year}
  {2020})}\BibitemShut {NoStop}%
\bibitem [{\citenamefont {Liu}\ \emph {et~al.}(2022)\citenamefont {Liu},
  \citenamefont {Zhang}, \citenamefont {Mann},\ and\ \citenamefont
  {Yu}}]{Liu:2021dnl}%
  \BibitemOpen
  \bibfield  {author} {\bibinfo {author} {\bibfnamefont {Z.}~\bibnamefont
  {Liu}}, \bibinfo {author} {\bibfnamefont {J.}~\bibnamefont {Zhang}}, \bibinfo
  {author} {\bibfnamefont {R.~B.}\ \bibnamefont {Mann}},\ and\ \bibinfo
  {author} {\bibfnamefont {H.}~\bibnamefont {Yu}},\ }\bibfield  {title}
  {\bibinfo {title} {{Does acceleration assist entanglement harvesting?}},\
  }\href {https://doi.org/10.1103/PhysRevD.105.085012} {\bibfield  {journal}
  {\bibinfo  {journal} {Phys. Rev. D}\ }\textbf {\bibinfo {volume} {105}},\
  \bibinfo {pages} {085012} (\bibinfo {year} {2022})},\ \Eprint
  {https://arxiv.org/abs/2111.04392} {arXiv:2111.04392 [quant-ph]} \BibitemShut
  {NoStop}%
\bibitem [{\citenamefont {Tjoa}\ and\ \citenamefont
  {Mann}(2020)}]{Tjoa2020vaidya}%
  \BibitemOpen
  \bibfield  {author} {\bibinfo {author} {\bibfnamefont {E.}~\bibnamefont
  {Tjoa}}\ and\ \bibinfo {author} {\bibfnamefont {R.~B.}\ \bibnamefont
  {Mann}},\ }\bibfield  {title} {\bibinfo {title} {{Harvesting correlations in
  Schwarzschild and collapsing shell spacetimes}},\ }\href
  {https://doi.org/https://doi.org/10.1007/JHEP08(2020)155} {\bibfield
  {journal} {\bibinfo  {journal} {J. High Energy Phys.}\ }\textbf {\bibinfo
  {volume} {08}}\bibinfo  {number} { (2020)},\ \bibinfo {pages}
  {155}}\BibitemShut {NoStop}%
\bibitem [{\citenamefont {Campos}\ and\ \citenamefont
  {Dappiaggi}(2021)}]{Campos-hyperbolicBH}%
  \BibitemOpen
\bibfield  {number} {  }\bibfield  {author} {\bibinfo {author} {\bibfnamefont
  {L.~d.~S.}\ \bibnamefont {Campos}}\ and\ \bibinfo {author} {\bibfnamefont
  {C.}~\bibnamefont {Dappiaggi}},\ }\bibfield  {title} {\bibinfo {title}
  {Ground and thermal states for the klein-gordon field on a massless
  hyperbolic black hole with applications to the anti-hawking effect},\ }\href
  {https://doi.org/10.1103/PhysRevD.103.025021} {\bibfield  {journal} {\bibinfo
   {journal} {Phys. Rev. D}\ }\textbf {\bibinfo {volume} {103}},\ \bibinfo
  {pages} {025021} (\bibinfo {year} {2021})}\BibitemShut {NoStop}%
\bibitem [{\citenamefont {Henderson}\ \emph {et~al.}(2022)\citenamefont
  {Henderson}, \citenamefont {Ding},\ and\ \citenamefont
  {Mann}}]{Henderson:2022oyd}%
  \BibitemOpen
  \bibfield  {author} {\bibinfo {author} {\bibfnamefont {L.~J.}\ \bibnamefont
  {Henderson}}, \bibinfo {author} {\bibfnamefont {S.~Y.}\ \bibnamefont
  {Ding}},\ and\ \bibinfo {author} {\bibfnamefont {R.~B.}\ \bibnamefont
  {Mann}},\ }\bibfield  {title} {\bibinfo {title} {{Entanglement harvesting
  with a twist}},\ }\href {https://doi.org/10.1116/5.0078314} {\bibfield
  {journal} {\bibinfo  {journal} {AVS Quantum Sci.}\ }\textbf {\bibinfo
  {volume} {4}},\ \bibinfo {pages} {014402} (\bibinfo {year} {2022})},\ \Eprint
  {https://arxiv.org/abs/2201.11130} {arXiv:2201.11130 [quant-ph]} \BibitemShut
  {NoStop}%
\bibitem [{\citenamefont {Robbins}\ \emph {et~al.}(2022)\citenamefont
  {Robbins}, \citenamefont {Henderson},\ and\ \citenamefont
  {Mann}}]{robbins2020entanglement}%
  \BibitemOpen
  \bibfield  {author} {\bibinfo {author} {\bibfnamefont {M.~P.~G.}\
  \bibnamefont {Robbins}}, \bibinfo {author} {\bibfnamefont {L.~J.}\
  \bibnamefont {Henderson}},\ and\ \bibinfo {author} {\bibfnamefont {R.~B.}\
  \bibnamefont {Mann}},\ }\bibfield  {title} {\bibinfo {title} {Entanglement
  amplification from rotating black holes},\ }\href
  {https://doi.org/10.1088/1361-6382/ac08a8} {\bibfield  {journal} {\bibinfo
  {journal} {Classical Quantum Gravity}\ }\textbf {\bibinfo {volume} {39}},\
  \bibinfo {pages} {02LT01} (\bibinfo {year} {2022})}\BibitemShut {NoStop}%
\bibitem [{\citenamefont {Gallock-Yoshimura}\ \emph {et~al.}(2021)\citenamefont
  {Gallock-Yoshimura}, \citenamefont {Tjoa},\ and\ \citenamefont
  {Mann}}]{Ken.Freefall.PhysRevD.104.025001}%
  \BibitemOpen
  \bibfield  {author} {\bibinfo {author} {\bibfnamefont {K.}~\bibnamefont
  {Gallock-Yoshimura}}, \bibinfo {author} {\bibfnamefont {E.}~\bibnamefont
  {Tjoa}},\ and\ \bibinfo {author} {\bibfnamefont {R.~B.}\ \bibnamefont
  {Mann}},\ }\bibfield  {title} {\bibinfo {title} {Harvesting entanglement with
  detectors freely falling into a black hole},\ }\href
  {https://doi.org/10.1103/PhysRevD.104.025001} {\bibfield  {journal} {\bibinfo
   {journal} {Phys. Rev. D}\ }\textbf {\bibinfo {volume} {104}},\ \bibinfo
  {pages} {025001} (\bibinfo {year} {2021})}\BibitemShut {NoStop}%
\bibitem [{\citenamefont {Simidzija}\ and\ \citenamefont
  {Mart\'{\i}n-Mart\'{\i}nez}(2018)}]{simidzija2018harvesting}%
  \BibitemOpen
  \bibfield  {author} {\bibinfo {author} {\bibfnamefont {P.}~\bibnamefont
  {Simidzija}}\ and\ \bibinfo {author} {\bibfnamefont {E.}~\bibnamefont
  {Mart\'{\i}n-Mart\'{\i}nez}},\ }\bibfield  {title} {\bibinfo {title}
  {Harvesting correlations from thermal and squeezed coherent states},\ }\href
  {https://doi.org/10.1103/PhysRevD.98.085007} {\bibfield  {journal} {\bibinfo
  {journal} {Phys. Rev. D}\ }\textbf {\bibinfo {volume} {98}},\ \bibinfo
  {pages} {085007} (\bibinfo {year} {2018})}\BibitemShut {NoStop}%
\bibitem [{\citenamefont {Gallock-Yoshimura}\ and\ \citenamefont
  {Mann}(2021)}]{GallockEntangledDetectors}%
  \BibitemOpen
  \bibfield  {author} {\bibinfo {author} {\bibfnamefont {K.}~\bibnamefont
  {Gallock-Yoshimura}}\ and\ \bibinfo {author} {\bibfnamefont {R.~B.}\
  \bibnamefont {Mann}},\ }\bibfield  {title} {\bibinfo {title} {Entangled
  detectors nonperturbatively harvest mutual information},\ }\href
  {https://doi.org/10.1103/PhysRevD.104.125017} {\bibfield  {journal} {\bibinfo
   {journal} {Phys. Rev. D}\ }\textbf {\bibinfo {volume} {104}},\ \bibinfo
  {pages} {125017} (\bibinfo {year} {2021})}\BibitemShut {NoStop}%
\bibitem [{\citenamefont {Sahu}\ \emph {et~al.}(2022)\citenamefont {Sahu},
  \citenamefont {Melgarejo-Lermas},\ and\ \citenamefont
  {Mart\'{\i}n-Mart\'{\i}nez}}]{SahuSabotage}%
  \BibitemOpen
  \bibfield  {author} {\bibinfo {author} {\bibfnamefont {A.}~\bibnamefont
  {Sahu}}, \bibinfo {author} {\bibfnamefont {I.}~\bibnamefont
  {Melgarejo-Lermas}},\ and\ \bibinfo {author} {\bibfnamefont {E.}~\bibnamefont
  {Mart\'{\i}n-Mart\'{\i}nez}},\ }\bibfield  {title} {\bibinfo {title}
  {Sabotaging the harvesting of correlations from quantum fields},\ }\href
  {https://doi.org/10.1103/PhysRevD.105.065011} {\bibfield  {journal} {\bibinfo
   {journal} {Phys. Rev. D}\ }\textbf {\bibinfo {volume} {105}},\ \bibinfo
  {pages} {065011} (\bibinfo {year} {2022})}\BibitemShut {NoStop}%
\bibitem [{\citenamefont {Tjoa}\ and\ \citenamefont
  {Mart\'{\i}n-Mart\'{\i}nez}(2021)}]{TjoaSignal}%
  \BibitemOpen
  \bibfield  {author} {\bibinfo {author} {\bibfnamefont {E.}~\bibnamefont
  {Tjoa}}\ and\ \bibinfo {author} {\bibfnamefont {E.}~\bibnamefont
  {Mart\'{\i}n-Mart\'{\i}nez}},\ }\bibfield  {title} {\bibinfo {title} {When
  entanglement harvesting is not really harvesting},\ }\href
  {https://doi.org/10.1103/PhysRevD.104.125005} {\bibfield  {journal} {\bibinfo
   {journal} {Phys. Rev. D}\ }\textbf {\bibinfo {volume} {104}},\ \bibinfo
  {pages} {125005} (\bibinfo {year} {2021})}\BibitemShut {NoStop}%
\bibitem [{\citenamefont {Ba\~nados}\ \emph {et~al.}(1992)\citenamefont
  {Ba\~nados}, \citenamefont {Teitelboim},\ and\ \citenamefont
  {Zanelli}}]{BTZ1}%
  \BibitemOpen
  \bibfield  {author} {\bibinfo {author} {\bibfnamefont {M.}~\bibnamefont
  {Ba\~nados}}, \bibinfo {author} {\bibfnamefont {C.}~\bibnamefont
  {Teitelboim}},\ and\ \bibinfo {author} {\bibfnamefont {J.}~\bibnamefont
  {Zanelli}},\ }\bibfield  {title} {\bibinfo {title} {Black hole in
  three-dimensional spacetime},\ }\href
  {https://doi.org/10.1103/PhysRevLett.69.1849} {\bibfield  {journal} {\bibinfo
   {journal} {Phys. Rev. Lett.}\ }\textbf {\bibinfo {volume} {69}},\ \bibinfo
  {pages} {1849} (\bibinfo {year} {1992})}\BibitemShut {NoStop}%
\bibitem [{\citenamefont {Ba\~nados}\ \emph {et~al.}(1993)\citenamefont
  {Ba\~nados}, \citenamefont {Henneaux}, \citenamefont {Teitelboim},\ and\
  \citenamefont {Zanelli}}]{BTZ2}%
  \BibitemOpen
  \bibfield  {author} {\bibinfo {author} {\bibfnamefont {M.}~\bibnamefont
  {Ba\~nados}}, \bibinfo {author} {\bibfnamefont {M.}~\bibnamefont {Henneaux}},
  \bibinfo {author} {\bibfnamefont {C.}~\bibnamefont {Teitelboim}},\ and\
  \bibinfo {author} {\bibfnamefont {J.}~\bibnamefont {Zanelli}},\ }\bibfield
  {title} {\bibinfo {title} {Geometry of the 2+1 black hole},\ }\href
  {https://doi.org/10.1103/PhysRevD.48.1506} {\bibfield  {journal} {\bibinfo
  {journal} {Phys. Rev. D}\ }\textbf {\bibinfo {volume} {48}},\ \bibinfo
  {pages} {1506} (\bibinfo {year} {1993})}\BibitemShut {NoStop}%
\bibitem [{\citenamefont {Lifschytz}\ and\ \citenamefont
  {Ortiz}(1994)}]{LifschytzBTZ}%
  \BibitemOpen
  \bibfield  {author} {\bibinfo {author} {\bibfnamefont {G.}~\bibnamefont
  {Lifschytz}}\ and\ \bibinfo {author} {\bibfnamefont {M.}~\bibnamefont
  {Ortiz}},\ }\bibfield  {title} {\bibinfo {title} {Scalar field quantization
  on the (2+1)-dimensional black hole background},\ }\href
  {https://doi.org/10.1103/PhysRevD.49.1929} {\bibfield  {journal} {\bibinfo
  {journal} {Phys. Rev. D}\ }\textbf {\bibinfo {volume} {49}},\ \bibinfo
  {pages} {1929} (\bibinfo {year} {1994})}\BibitemShut {NoStop}%
\bibitem [{\citenamefont {Henderson}\ \emph {et~al.}(2020)\citenamefont
  {Henderson}, \citenamefont {Hennigar}, \citenamefont {Mann}, \citenamefont
  {Smith},\ and\ \citenamefont {Zhang}}]{Henderson2019anti-hawking}%
  \BibitemOpen
  \bibfield  {author} {\bibinfo {author} {\bibfnamefont {L.~J.}\ \bibnamefont
  {Henderson}}, \bibinfo {author} {\bibfnamefont {R.~A.}\ \bibnamefont
  {Hennigar}}, \bibinfo {author} {\bibfnamefont {R.~B.}\ \bibnamefont {Mann}},
  \bibinfo {author} {\bibfnamefont {A.~R.}\ \bibnamefont {Smith}},\ and\
  \bibinfo {author} {\bibfnamefont {J.}~\bibnamefont {Zhang}},\ }\bibfield
  {title} {\bibinfo {title} {Anti-hawking phenomena},\ }\href
  {https://doi.org/https://doi.org/10.1016/j.physletb.2020.135732} {\bibfield
  {journal} {\bibinfo  {journal} {Physics Letters B}\ }\textbf {\bibinfo
  {volume} {809}},\ \bibinfo {pages} {135732} (\bibinfo {year}
  {2020})}\BibitemShut {NoStop}%
\bibitem [{\citenamefont {de~Souza~Campos}\ and\ \citenamefont
  {Dappiaggi}(2021)}]{Campos-RobinBC}%
  \BibitemOpen
  \bibfield  {author} {\bibinfo {author} {\bibfnamefont {L.}~\bibnamefont
  {de~Souza~Campos}}\ and\ \bibinfo {author} {\bibfnamefont {C.}~\bibnamefont
  {Dappiaggi}},\ }\bibfield  {title} {\bibinfo {title} {The anti-hawking effect
  on a btz black hole with robin boundary conditions},\ }\href
  {https://doi.org/https://doi.org/10.1016/j.physletb.2021.136198} {\bibfield
  {journal} {\bibinfo  {journal} {Physics Letters B}\ }\textbf {\bibinfo
  {volume} {816}},\ \bibinfo {pages} {136198} (\bibinfo {year}
  {2021})}\BibitemShut {NoStop}%
\bibitem [{\citenamefont {Robbins}\ and\ \citenamefont
  {Mann}(2021)}]{Robbins-Anti-Hawking}%
  \BibitemOpen
  \bibfield  {author} {\bibinfo {author} {\bibfnamefont {M.~P.~G.}\
  \bibnamefont {Robbins}}\ and\ \bibinfo {author} {\bibfnamefont {R.~B.}\
  \bibnamefont {Mann}},\ }\href {https://doi.org/10.48550/ARXIV.2107.01648}
  {\bibinfo {title} {Anti-hawking phenomena around a rotating btz black hole}}
  (\bibinfo {year} {2021})\BibitemShut {NoStop}%
\bibitem [{\citenamefont {Jennings}(2010)}]{Jennings:2010vk}%
  \BibitemOpen
  \bibfield  {author} {\bibinfo {author} {\bibfnamefont {D.}~\bibnamefont
  {Jennings}},\ }\bibfield  {title} {\bibinfo {title} {{On the response of a
  particle detector in Anti-de Sitter spacetime}},\ }\href
  {https://doi.org/10.1088/0264-9381/27/20/205005} {\bibfield  {journal}
  {\bibinfo  {journal} {Class. Quant. Grav.}\ }\textbf {\bibinfo {volume}
  {27}},\ \bibinfo {pages} {205005} (\bibinfo {year} {2010})},\ \Eprint
  {https://arxiv.org/abs/1008.2165} {arXiv:1008.2165 [gr-qc]} \BibitemShut
  {NoStop}%
\bibitem [{\citenamefont {Mart\'{\i}n-Mart\'{\i}nez}\ and\ \citenamefont
  {Rodriguez-Lopez}(2018)}]{EMM.Relativistic.quantum.optics}%
  \BibitemOpen
  \bibfield  {author} {\bibinfo {author} {\bibfnamefont {E.}~\bibnamefont
  {Mart\'{\i}n-Mart\'{\i}nez}}\ and\ \bibinfo {author} {\bibfnamefont
  {P.}~\bibnamefont {Rodriguez-Lopez}},\ }\bibfield  {title} {\bibinfo {title}
  {Relativistic quantum optics: The relativistic invariance of the light-matter
  interaction models},\ }\href {https://doi.org/10.1103/PhysRevD.97.105026}
  {\bibfield  {journal} {\bibinfo  {journal} {Phys. Rev. D}\ }\textbf {\bibinfo
  {volume} {97}},\ \bibinfo {pages} {105026} (\bibinfo {year}
  {2018})}\BibitemShut {NoStop}%
\bibitem [{\citenamefont {Mart\'{\i}n-Mart\'{\i}nez}\ \emph
  {et~al.}(2020)\citenamefont {Mart\'{\i}n-Mart\'{\i}nez}, \citenamefont
  {Perche},\ and\ \citenamefont {de~S.~L.~Torres}}]{Tales2020GRQO}%
  \BibitemOpen
  \bibfield  {author} {\bibinfo {author} {\bibfnamefont {E.}~\bibnamefont
  {Mart\'{\i}n-Mart\'{\i}nez}}, \bibinfo {author} {\bibfnamefont {T.~R.}\
  \bibnamefont {Perche}},\ and\ \bibinfo {author} {\bibfnamefont
  {B.}~\bibnamefont {de~S.~L.~Torres}},\ }\bibfield  {title} {\bibinfo {title}
  {General relativistic quantum optics: Finite-size particle detector models in
  curved spacetimes},\ }\href {https://doi.org/10.1103/PhysRevD.101.045017}
  {\bibfield  {journal} {\bibinfo  {journal} {Phys. Rev. D}\ }\textbf {\bibinfo
  {volume} {101}},\ \bibinfo {pages} {045017} (\bibinfo {year}
  {2020})}\BibitemShut {NoStop}%
\bibitem [{\citenamefont {Conroy}\ and\ \citenamefont
  {Taylor}(2022)}]{Conroy-extremealBH}%
  \BibitemOpen
  \bibfield  {author} {\bibinfo {author} {\bibfnamefont {A.}~\bibnamefont
  {Conroy}}\ and\ \bibinfo {author} {\bibfnamefont {P.}~\bibnamefont
  {Taylor}},\ }\bibfield  {title} {\bibinfo {title} {Response of an
  unruh-dewitt detector near an extremal black hole},\ }\href
  {https://doi.org/10.1103/PhysRevD.105.085001} {\bibfield  {journal} {\bibinfo
   {journal} {Phys. Rev. D}\ }\textbf {\bibinfo {volume} {105}},\ \bibinfo
  {pages} {085001} (\bibinfo {year} {2022})}\BibitemShut {NoStop}%
\bibitem [{\citenamefont {Brenna}\ \emph {et~al.}(2016)\citenamefont {Brenna},
  \citenamefont {Mann},\ and\ \citenamefont
  {Martín-Martínez}}]{Brenna2016anti-unruh}%
  \BibitemOpen
  \bibfield  {author} {\bibinfo {author} {\bibfnamefont {W.}~\bibnamefont
  {Brenna}}, \bibinfo {author} {\bibfnamefont {R.~B.}\ \bibnamefont {Mann}},\
  and\ \bibinfo {author} {\bibfnamefont {E.}~\bibnamefont
  {Martín-Martínez}},\ }\bibfield  {title} {\bibinfo {title} {{Anti-Unruh
  phenomena}},\ }\href
  {https://doi.org/https://doi.org/10.1016/j.physletb.2016.04.002} {\bibfield
  {journal} {\bibinfo  {journal} {Physics Letters B}\ }\textbf {\bibinfo
  {volume} {757}},\ \bibinfo {pages} {307 } (\bibinfo {year}
  {2016})}\BibitemShut {NoStop}%
\bibitem [{\citenamefont {Garay}\ \emph {et~al.}(2016)\citenamefont {Garay},
  \citenamefont {Mart\'{\i}n-Mart\'{\i}nez},\ and\ \citenamefont
  {de~Ram\'on}}]{Garay2016anti-unruh}%
  \BibitemOpen
  \bibfield  {author} {\bibinfo {author} {\bibfnamefont {L.~J.}\ \bibnamefont
  {Garay}}, \bibinfo {author} {\bibfnamefont {E.}~\bibnamefont
  {Mart\'{\i}n-Mart\'{\i}nez}},\ and\ \bibinfo {author} {\bibfnamefont
  {J.}~\bibnamefont {de~Ram\'on}},\ }\bibfield  {title} {\bibinfo {title}
  {{Thermalization of particle detectors: The Unruh effect and its reverse}},\
  }\href {https://doi.org/10.1103/PhysRevD.94.104048} {\bibfield  {journal}
  {\bibinfo  {journal} {Phys. Rev. D}\ }\textbf {\bibinfo {volume} {94}},\
  \bibinfo {pages} {104048} (\bibinfo {year} {2016})}\BibitemShut {NoStop}%
\end{thebibliography}%

\end{document}